\documentclass[12pt,twoside]{article}
\usepackage[mathscr]{eucal}
\usepackage{amsmath,amsfonts,amssymb,amsthm}
\usepackage{hyperref}
\usepackage[matrix,arrow]{xy}
\usepackage{times}

\advance\textheight\topskip
\def\BGST{Barnich:2004cr}
\def\BGL{Batalin:2001je}
\def\GL{Grigoriev:2000rn}

\renewcommand{\r}{{(r)}}

\def\unf{{\rm unf}}
\def\sd{S^\dagger}
\def\bsd{\bar S^\dagger}
\def\ss{\bar s}
\def\nn{\bar n}
\def\ww{w}
\def\lift{\mathcal{K}}

\newcommand{\Pj}{\mathscr{{P}}}

\renewcommand{\tilde}{\widetilde}
\renewcommand{\hat}{\widehat}
\newtheorem{prop}{Proposition}[section]
\newtheorem{lemma}[prop]{Lemma}

\renewcommand{\simeq}{\cong}
\newcommand{\ul}[1]{{\underline{#1}}}

\newcommand{\T}{\mathrm{T}}

\newcommand{\bref}[1]{\textbf{\ref{#1}}}

\newcommand{\dmn}{d}  % dimension of the space--time

\newcommand{\Ker}{\mathop{\mathrm{Ker}}}
\newcommand{\im}{\mathop{\mathrm{Im}}}

\newcommand{\p}[1]{|#1|}
\newcommand{\gh}[1]{\mathrm{gh}(#1)}

\newcommand{\map}{\,\mathrm{:}\,}

\newcommand{\dd}{\partial}
\renewcommand{\d}{\partial}

\newcommand{\tensor}{\otimes}

\renewcommand{\geq}{\,{\geqslant}\,}

\newcommand{\inner}[2]{\langle #1{,}\,#2\rangle}
\newcommand{\binner}[2]{%
  {\langle}\kern-4.15pt{\langle}#1{,}\,#2{\rangle}\kern-4.15pt{\rangle}}
\newcommand{\commut}[2]{[#1{,}\,#2]}

\newcommand{\pb}[2]{\left\{{}#1{},{}#2{}\right\}}
\newcommand{\ab}[2]{\big(#1,#2\big)}

\newcommand{\half}{\mathchoice{%
    \ffrac{1}{2}}{\frac{1}{2}}{\frac{1}{2}}{\frac{1}{2}}}

\newcommand{\ffrac}[2]{\raisebox{.5pt}%
  {\footnotesize$\displaystyle\frac{#1}{#2}$}\kern1pt}

\newcommand{\brst}{\mathsf{\Omega}}
\newcommand{\red}{\mathrm{red}}

\newcommand{\st}[2]{\overset{#1}{#2}}

\newcommand{\dl}[1]{\mathchoice{\ffrac{\dd}{\dd #1}}{\frac{\dd}{\dd
      #1}}{\ffrac{\dd}{\dd #1}}{\ffrac{\dd}{\dd #1}}}

\newcommand{\ddl}[2]{\ffrac{\dd #1}{\dd #2}}

\newcommand{\vac}{|0\rangle}

\newcommand{\fR}{\mathbb{R}}

\newcommand{\bundle}{\boldsymbol}

\newcommand{\derham}{\boldsymbol{d}}

\newcommand{\manifold}[1]{\mathscr{#1}}
\newcommand{\manX}{\manifold{X}}

\def\cD{\mathcal{D}}
\def\cE{\mathcal{E}}
\def\cF{\mathcal{F}}
\def\cG{\mathcal{G}}
\def\cH{\mathcal{H}}

\def\cL{\mathcal{L}}

\def\cP{\mathcal{P}}

\def\cS{\mathcal{S}}
\def\cT{\mathcal{T}}

\def\cV{\mathcal{V}}

\numberwithin{equation}{section} \makeatletter
\@addtoreset{equation}{section}

\begin{document}

\def\mytitle{Parent form for higher spin fields on anti-de Sitter
  space}

\pagestyle{myheadings}
\markboth{\textsc{\small Barnich, Grigoriev}}{%
  \textsc{\small \mytitle}}
\addtolength{\headsep}{4pt}

\begin{flushright}\small
ULB-TH/06-01\\
FIAN-TD/05-06\\
\texttt{hep-th/0602166}\\[-3pt]
\end{flushright}

\begin{centering}

  \vspace{1cm}

  \textbf{\Large{\mytitle}}

\vspace{1cm}

  \vspace{1.5cm}

  {\large Glenn Barnich$^{a,*}$ and Maxim Grigoriev$^{b}$ }

\vspace{1.5cm}

\begin{minipage}{.9\textwidth}\small \it \begin{center}
   $^a$Physique Th\'eorique et Math\'ematique, Universit\'e Libre de
   Bruxelles\\ and \\ International Solvay Institutes, \\ Campus
   Plaine C.P. 231, B-1050 Bruxelles, Belgium \end{center}
\end{minipage}

    \vspace{.5cm}

\begin{minipage}{.9\textwidth}\small \it \begin{center}
   $^b$Tamm Theory Department, Lebedev Physics
   Institute,\\ Leninsky prospect 53, 119991 Moscow, Russia\end{center}
\end{minipage}

\end{centering}

\vspace{1cm}

\begin{center}
  \begin{minipage}{.9\textwidth}
    \textsc{Abstract}.  We construct a first order parent field theory
    for free higher spin gauge fields on constant curvature spaces. As
    in the previously considered flat case, both the original
    formulation by Fronsdal and the unfolded one by Vasiliev can be
    reached by two different straightforward reductions. The parent
    theory itself is formulated using a higher dimensional embedding
    space. It turns out to be geometrically extremely transparent and
    free of the intricacies of both of its reductions.
  \end{minipage}
\end{center}

\vfill

\noindent
\mbox{}
\raisebox{-3\baselineskip}{%
  \parbox{\textwidth}{\mbox{}\hrulefill\\[-4pt]}}
{\scriptsize$^*$ Senior Research Associate of the National
  Fund for Scientific Research (Belgium).}

\thispagestyle{empty}
\newpage

\begin{small}
{\addtolength{\parskip}{-1.5pt}
 \tableofcontents}
\end{small}
\newpage
\section{Introduction}
\label{sec:introduction}

Progress in the subject of higher spin gauge fields has often been
related with the construction of new equivalent formulations of the
theory. A Lagrangian formulation at the free
level~\cite{Fierz:1939ix,Singh:1974qz,Fronsdal:1978rb,Fronsdal:1979vb}
required the introduction of a carefully selected set of auxiliary
fields. Other auxiliary fields were needed for the BRST reformulation,
directly inspired by string field theory (see
e.g~\cite{Siegel:1984ap,%
  Siegel:1984wx,Siegel:1984xd,Siegel:1985tw,Neveu:1985ya,Neveu:1985cx,%
  Witten:1985cc,Ohta:1985zw,Thorn:1989hm} and references therein), in
terms of the field theory associated to a BRST first quantized
particle model~\cite{Ouvry:1986dv,Bengtsson:1986ys,Henneaux:1987cp}
(see e.g.~\cite{Bouatta:2004kk} for a review). This reformulation
explicitly revealed the relation with the tensionless limit of
string theory and provided a compact Lagrangian description at
the free level. Similarly, finding a consistent
interaction~\cite{Vasiliev:2003ev,Vasiliev:1990en} on an anti-de
Sitter (AdS) background was done exclusively in the \textit{unfolded
  formulation}~\cite{Vasiliev:1988xc,Vasiliev:1988sa} that also
provides a natural framework for various other problems of higher spin
theories~\cite{Vasiliev:2001zy,Alkalaev:2003qv,Sezgin:2001ij}.

Further developments, such as for example understanding whether the
recently constructed interaction in an AdS background admits a
Lagrangian formulation, require a good control over various equivalent
formulations differing by auxiliary and pure gauge fields. As a first
step, one would like to explicitly relate the unfolded formulation and
the BRST formulation. In the case of a flat background, this has been
done recently through the construction of a parent theory~\cite{\BGST}
from which both formulations can be reached through consistent
reductions. Furthermore, some algebraic structures that are hidden in
the BRST or unfolded formulations appear more transparently in the
parent theory or some of its intermediate reductions. The objective of
the present paper is to extend these results to an AdS background.

From a more technical point of view, understanding free higher spin
gauge fields in terms of a first-quantized generally covariant
particle model has two advantages: firstly, it allows one to transpose
the arsenal of cohomological methods available at the BRST first
quantized level to the gauge field theory. In particular, because
auxiliary fields and pure gauge degrees of freedom can be identified
with cohomologically trivial pairs at the first-quantized level,
showing the equivalence of various formulations boils down to a
straightforward exercise in homological algebra. Secondly, well known
quantization techniques for complicated constrained systems in curved
spaces can be used to construct, new, more transparent descriptions of
the gauge field theory.

In this paper, we combine a set of ideas available in the literature
to construct the parent theory of higher spin gauge fields on AdS: the
use of an embedding space with vielbeins and connections
\cite{Vasiliev:2001wa,Vasiliev:2003ev,MacDowell:1977jt,Stelle:1979aj}
(see also \cite{Bekaert:2005vh} for a review) and a Fedosov-type
approach for constrained systems
\cite{Fedosov-book,Bordemann:1997er,Batalin:1989mb,\GL,\BGL} in order
to achieve a generally covariant description. The resulting parent
theory is completely natural from a geometrical point of view and
admits a transparent algebraic structure with the simplest possible
numerical factors. We first show how it reduces to the BRST based
``metric-like'' description of higher spins on AdS
\cite{Bengtsson:1990un,Buchbinder:2001bs,Sagnotti:2003qa} that is
directly related to Fronsdal's original formulation and then analyze
the reduction to Vasiliev's unfolded formulation
\cite{Lopatin:1988hz,Vasiliev:2001wa}. Along the way, we construct
various new intermediate descriptions with less variables but more
complicated structure. We hope that the parent theory or one of its
intermediate reductions will be useful for resolving the
above-mentioned problem of compatibility between Lagrangian and
interaction.

The paper is organized as follows: in the next section, we briefly
recall how to associate a gauge field theory to a BRST first-quantized
system.  We also discuss some reduction techniques on the first
quantized level and the relation with generalized auxiliary
fields. Finally, we comment on the existence of Lagrangians associated
with BRST field theories. In Section \bref{sec:particle}, we give the
details on the embedding and the covariantization procedure by
constructing the parent theory for a scalar particle on AdS. We also
discuss its reductions to standard and unfolded form. The inclusion of
additional internal degrees of freedom to get our main result, the
parent theory for higher spin gauge fields on AdS, is then
straightforward and done in Section \bref{sec:fronsdal}. The explicit
reductions are more involved. We summarize the main steps in the rest
of Section~\bref{sec:fronsdal}. Mathematical and technical details on
reductions are relegated to the appendix.

\section{Gauge field theories associated to first-quantized
  systems}

\subsection{BRST differential and equations of motion}
\label{sec:2.1}

Let us briefly recall some basic facts about free field theories
associated to BRST first-quantized systems with vanishing Hamiltonian.
More detailed expositions can be found for instance
in~\cite{Gaberdiel:1997ia,Barnich:2003wj} and in \cite{\BGST}, which
we follow here.

Suppose we are given with a quantum BRST system whose space of states
is the space of sections $\Gamma(\bundle{\cH})$ of a vector bundle
$\bundle\cH$ over a space-time manifold $\manX$. Locally, the space of
states can be identified with functions on $\manX$ taking values in
the graded superspace $\cH$. The degree is identified with the ghost
number and denoted by $\gh{\cdot}$. The BRST operator $\brst\map
\Gamma(\bundle{\cH}) \to \Gamma(\bundle{\cH})$, $\gh{\brst}=1$ is
assumed to be a Grassmann odd differential operator of finite
order. Locally, it is a differential operator with coefficients in
linear operators on $\cH$. In what follows such a BRST system is
referred to as a BRST first-quantized system
$(\brst,\Gamma(\bundle{\cH}))$.

A local gauge field theory is associated to
$(\brst,\Gamma(\bundle{\cH}))$ in the following way. If $e_A$ is a
real frame of the bundle $\bundle{\cH}$, a generic section is given by
$\bundle\phi={\bundle\phi}^A(x)e_A$. One then introduces an
independent field $\psi^A(x)$ for each component, with Grassmann
parity and the ghost number prescribed by $\p{\psi^A}=\p{e_A}$ and
$\gh{\psi^A}=-\gh{e_A}$.  All component fields are combined into a
single string field
\begin{equation}
  \Psi=e_A \tensor \psi^A\,,
\end{equation}
understood (locally) as an element of the tensor product of $\cH$ and
the algebra of local functions, i.e., functions in $\psi^A$ and their
space-time derivatives, see~\cite{\BGST} for details. The field theory
associated with the first-quantized system
$(\brst,\Gamma(\bundle{\cH}))$ is determined by the BRST differential
$s\Psi=\brst\Psi$ or, in terms of components, $s\psi^A=\brst^A_B
\psi^B$. Differential $s$ is extended to arbitrary local functions by
requiring that $s$ satisfies the Leibnitz rule and commutes with the
total derivative
\begin{equation}\label{d-total}
  \dd_\mu=\dl{x^\mu}+\psi^A_{,\mu}\dl{\psi^A}+
  \psi^A_{,\mu\nu}\dl{\psi^A_{,\nu}}+\ldots.
\end{equation}
In particular, the equations of motion have the form $s\Psi^{(-1)}=0
\Leftrightarrow \brst \Psi^{(0)}=0$ while the gauge transformations
are identified with $\delta\Psi^{(0)}=\brst\Psi^{(1)}$ where
ghost-number-one fields in $\Psi^{(1)}$ are replaced with gauge
parameters. Here and in what follows we use the decomposition
$\Psi=\sum_n\Psi^{(n)}$ of a string field into components $\Psi^{(n)}$
containing fields at ghost number $n$.

\subsection{Reductions}
\label{sec:reductions}

Consider a not necessarily linear nor Lagrangian BRST gauge field
theory described by a differential $s$, understood as a vector field
on the space of fields $\psi^A$ and their derivatives.  The
differential $s$ is assumed to be local, i.e., $s\psi^A$ involve
derivatives of finite order, and to be commuting with the total
derivative $\d_\mu$. Even in this more general non linear context, it
is still useful to combine all the fields into a string field
$\Psi$. The equations of motion for the physical fields are then given
by $s\Psi^{(-1)}\big|_{\Psi^{k}=0,\,k\neq 0}=0$ while the gauge
symmetries are determined by
$\delta\Psi^{(0)}=s\Psi^{(0)}|_{\Psi^{(k)}=0,\,k\neq 0,1}$ with
ghost-number-$1$ component fields of $\Psi^{(1)}$ replaced by gauge
parameters. 

Suppose that, after an invertible change of coordinates involving
derivatives if necessary, the set of fields $\psi^A$ splits into
$\varphi^\alpha,w^a,v^a$ such that equations $sw^a|_{w^a=0}=0$
(understood as algebraic equations in the space of fields and their
derivatives) are equivalent to $v^a=V^a[\varphi^\alpha]$, i.e., can be
algebraically solved for fields $v^a$. One then says that fields $w,v$
are generalized auxiliary fields. The field theory described by $s$ is
then equivalent to that described by the reduced differential $\tilde
s$ acting on the space of fields $\varphi^\alpha$ and their
derivatives and defined by $\tilde s
\varphi^\alpha=s\varphi^\alpha|_{w^a=0,\, v^a=V^a[\varphi]}$ (see
\cite{\BGST} for more details). In the Lagrangian framework, fields
$w,v$ are in addition required to be second-class constraints in the
antibracket sense.  In this context, generalized auxiliary fields were
originally proposed in~\cite{Dresse:1990dj}.  Note that generalized
auxiliary fields comprise both standard auxiliary fields and pure
gauge degrees of freedom, together with associated ghost and
antifields.

In the case where the gauge field theory is a linear theory associated
with a BRST first-quantized system $(\brst,\Gamma(\bundle{\cH}))$, one
can proceed with the reductions at the first-quantized level. To
identify a first-quantized counterpart of elimination of generalized
auxiliary fields, we need to recall the notion of consistent reduction
of a first-quantized gauge system discussed in \cite{\BGST}.  In order
to do so, we use the concept of algebraic invertibility: a
differential operator $O \map \Gamma(\bundle\cH) \to
\Gamma(\bundle\cH)$ is algebraically invertible iff it is invertible
in the space of differential operators of (graded) finite order. In
terms of a local frame a differential operator has the form
$O=O^A_B(x,\dl{x})$ so that $O\bundle\phi=O(e_A{\bundle\phi}^A)=e_B
O^B_A {\bundle\phi}^A$ for $\bundle\phi\in \Gamma(\bundle\cH)$. Note that the
derivative-independent part of an algebraically invertible operator is
an invertible matrix.
\begin{prop}\label{prop:red-0}
  Let $\bundle{\cH}$ decompose into a direct sum of vector bundles
  $\bundle\cH=\bundle\cE\oplus\bundle\cG\oplus\bundle\cF$ and the
  component
  $\st{\bundle\cG\bundle\cF}{\brst}=\Pj_{\bundle\cG}\brst\Pj_{\bundle\cF}$,
  with $\Pj_{\bundle\cG}$,$\Pj_{\bundle\cF}$ denoting the projector to
  $\Gamma(\bundle\cG)$, resp.  $\Gamma(\bundle\cF)$, be algebraically
  invertible as an operator from $\Gamma(\bundle\cF)$ to
  $\Gamma(\bundle\cG)$. Then the system $(\brst,\Gamma(\bundle{\cH}))$
  can be consistently reduced to $(\tilde\brst,\Gamma(\bundle{\cE}))$
  with
\begin{equation}\label{tilde-brst}
  \tilde\brst
  =(\st{\bundle\cE\bundle\cE}{\brst}
  - \st{\bundle\cE\bundle\cF}{\brst}(\st{\bundle\cG\bundle\cF}{\brst})^{-1}
  \st{\bundle\cG\bundle\cE}{\brst})\qquad\quad
\tilde\brst\map\Gamma(\bundle{\cE})\to\Gamma(\bundle{\cE})\,.
\end{equation}
In this case, the gauge field theories associated with
$(\brst,\Gamma(\bundle{\cH}))$ and
$(\tilde\brst,\Gamma(\bundle{\cE}))$ are related by elimination of
generalized auxiliary fields.
\end{prop}
In Appendix~\bref{sec:A1} we recall a useful proposition which allows
one to systematically study various consistent reductions and discuss
the relation with the so-called $\cD$-module approach to linear
partial differential equations.

To conclude this discussion of consistent reductions in first
quantized terms, let us note that this procedure controls the problem
of identifying generalized auxiliary fields in the non linear case as
well. Indeed, suppose that the non linear theory corresponds to a
consistent deformation of a linear theory associated to
$(\brst,\Gamma(\bundle\cH))$, i.e., the non linear BRST differential
has the form
\begin{equation}
  s=s_0+gs_1+g^2s_2+\ldots\,,\qquad s^2=0
\end{equation}
with $s_0\Psi=\brst\Psi$ the free BRST differential and $g$ a coupling
constant understood as formal deformation parameter. Now, if
the fields of the theory split into $w^a,v^a,\varphi^\alpha$ so that
$s_0w^a=0$ can be algebraically solved as $v^a=V^a_0[\varphi]$ at
$w_a=0$, i.e., if  $w,v$ are generalized auxiliary fields of the free
theory,  it is then easy to see that they are also generalized
auxiliary fields for the deformed theory. Namely, at $w=0$ equations
$sw^a=0$ can be algebraically solved as
\begin{equation}
  v^a=V_0^a[\varphi]+gV_1^a[\varphi]+\ldots\,,
\end{equation}
order by order in $g$. Note however that in this setting all
quantities, such as the reduced BRST differential for instance, are
formal power series in the deformation parameter $g$. In particular,
even if $s$ is polynomial, the reduced differential $\tilde s$ can be
an infinite series whose convergence is a separate question.

\subsection{Lagrangians}
\label{sec:Lagrangian}

Whenever there exists an inner product that makes the BRST operator
$\brst$ hermitian, the action that gives rise to the equations of
motion is
\begin{equation}\label{eq:physaction}
  \bundle{S}^{\mathrm{ph}}[\Psi^{(0)}]=-\half
  \inner{\Psi^{(0)}}{\brst \Psi^{(0)}}\,,
\end{equation}
while the functional
\begin{equation}\label{eq:psiopsi}
  \bundle{S}[\Psi]=-\half
\inner{\Psi}{\brst \Psi}
\end{equation}
is the Batalin-Vilkovisky master
action~\cite{Thorn:1987qj,Bochicchio:1987zj,Bochicchio:1987bd,
  Thorn:1989hm} associated with~\eqref{eq:physaction}.

As in the beginning of the previous subsection, consider a not
necessarily Lagrangian or linear BRST gauge field theory described by
a BRST differential $s$ and let us also assume that the set of fields
$\psi^A$ splits into fields $\varphi^\alpha$, $w^a$, and $v^a$ such that
$w^a$ and $v^a$ are generalized auxiliary fields. Let $\tilde s$ be
the reduced BRST differential acting on the space of fields
$\varphi^\alpha$ according to $\tilde s
\varphi^\alpha=s\varphi^\alpha|_{w^a=0,\, v^a=V^a[\varphi]}$.

Suppose now in addition that the reduced system described by $\tilde
s$ is Lagrangian which, on the level of the master action, is
expressed through the existence of an antibracket
$\ab{\cdot}{\cdot}_{\rm red}$ on the space of local functions in
$\varphi^\alpha,\d_{\cdot}\varphi_\alpha,\ldots$ such that $\tilde s$ is
generated by a master action $\tilde S[\varphi]$, i.e., $\tilde
s=\ab{\tilde S}{\cdot}_{\red}$. Note that under appropriate regularity
conditions, this is in fact equivalent to the existence of a standard
Lagrangian for the equations
$s\Psi^{(-1)}\big|_{\Psi^{(k)}=0,\,k\neq 0}=0$. Under these
assumptions, one can show that the original theory described by $s$
can be also made Lagrangian by introducing ``generalized'' Lagrange
multipliers.  Generalized Lagrange multipliers are related to ordinary
Lagrange multipliers in the same way as generalized auxiliary fields
are related to ordinary ones: they are Lagrange multiplies on the
level of the master action, instead of the classical action.

In order to see this, let us introduce adapted coordinates
$\varphi^\alpha,w^a,v^a=sw^a$ as new independent coordinates on the space
of fields. Moreover one can always redefine $\varphi^\alpha$ such that
$s\varphi^\alpha$ are functions only of $\varphi^\alpha$ and their
derivatives (see \cite{\BGST} for a proof). In the new coordinate
system the differential $s$ takes the form
\begin{equation}
  s=s^\alpha[\varphi]\dl{\varphi^\alpha}+v^a\dl{w^a} +\ldots\,,
\end{equation}
where dots denote the terms acting on derivatives.

The generalized Lagrange multipliers are then the new fields $v^*_a$
and $w^*_a$ with $\p{v^*_a}=\p{v^a}+1$, $\p{w^*_a}=\p{w^a}+1$ and
$\gh{v^*_a}=-\gh{v^a}-1$, $\gh{w^*_a}=-\gh{w^*_a}-1$.
The extended space is equipped with the following antibracket structure
\begin{equation}
  \ab{\varphi^\alpha}{\varphi_\beta}=\ab{\varphi^\alpha}{\varphi_\beta}_{\rm
    red}\,, \qquad 
\ab{v^a}{v^*_b}=\delta^a_b\,,\qquad \ab{w^a}{w^*_b}=\delta^a_b\,,
\end{equation}
with all the other basic antibrackets vanishing. The bracket is
extended in the standard way (see e.g. \cite{Barnich:1996mr}) to
general local functions such that it satisfies the Leibnitz rule for
the second argument and commutes with the total derivative acting on
the second argument. The master action that describes the Lagrangian
structure of the original theory is then given by
\begin{equation}
  S=\tilde S - \int d^dx \,v^a w^*_a\,.
\end{equation}
It obviously satisfies the master equations $\half\ab{S}{S}=0$.  If
$f$ does not depend on $v^*,w^*$ and their derivatives then
\begin{equation}
  \ab{S}{f}=sf\,,
\end{equation}
where $s$ is the original BRST differential.

Furthermore, fields $v,w,v^*,w^*$ are obviously generalized auxiliary
fields in the sense of~\cite{Dresse:1990dj}. Indeed, the equations of
motion obtained by varying with respect to fields $v$ and $w^*$ can be
algebraically solved for these variables. The reduced master action is
$\tilde S[\varphi]$ which establishes the equivalence of the extended
and the reduced theories as Lagrangian field theories.

\section{Scalar particle on AdS}\label{sec:particle}

\subsection{ (A)dS space as an embedding}

We take the standard approach and describe gauge systems on constant
curvature spaces by embedding the latter in a flat pseudo-Euclidean
space. More precisely, we consider the surface $\manX_0\subset \fR^{d+1}$
described by
\begin{equation}
  \eta_{AB}X^AX^B+l^2=0\,,
\end{equation}
where $X^A$, $A=0,\dots,\dmn$ stand for the standard coordinates in
$\fR^{d+1}$ while the metric is chosen as
$\eta_{AB}=diag(-1,1,\dots,1,-1)$. When $l^2>0$, the manifold
$\manX_0$ describes AdS space, the case $l^2<0$ corresponds to dS
space. In what follows we explicitly consider the case where $\manX_0$
is AdS space, but the analysis remains the same for other constant
curvature spaces.

The main advantage of an embedding space over an intrinsic description
is the transparent form of the isometries. Similarly, in the higher
spin gauge field context, the characterization of the ``vacuum
symmetries'', i.e., of the gauge transformations that leave the
background solution invariant, is considerably simplified when one
uses an embedding space (see e.g.~\cite{Bekaert:2005vh}). 

\subsection{BRST operator}

To demonstrate the approach in the most simple case, we consider the
quantum theory of a massless scalar particle. At the classical level
the phase space is just given by flat space with coordinates $X^A,P_A$
subjected to the standard Poisson bracket relations. The effective
phase space of a particle on $\manX_0$ is described by the
second class constraints
\begin{equation}
X^2+l^2=0\,, \qquad XP=0
\end{equation}
together with the mass-shell constraint $P^2=0$. In order to have a
description in terms of first class constraints only, the geometrical
constraint $X^2+l^2=0$ is excluded from the initial set of constraints
and treated as a partial gauge fixing condition (see
e.g.~\cite{Bonelli:2003zu}). Note that in terms of wave functions this
reproduces the well known approach of~\cite{Fronsdal:1979vb}. As a
result, one has the following set of first class constraints
\begin{equation}
\label{eq:GL-constr}
 L=P_A \eta^{AB} P_B\,, \qquad M=P_AX^A\,,
\end{equation}
which form a closed algebra $\pb{L}{M}=-2L$.

In principle, one can construct the quantum theory by treating $P,X$
as operators represented on functions in $X$ and build the associated
gauge field theory, which then describes a scalar field on $\manX_0$.
Indeed, in this representation the constraint $XP$ completely fixes
the radial dependence of wave functions which then can be considered 
as fields on $\manX_0$.
However, we now take a different route and first
extend the constrained system even further. What we want is an
explicitly covariant formulation of the system, in terms of bundles on
$\manX_0$ with fibers related to the embedding space. For this
purpose, we generalize the parent theory of~\cite{Barnich:2004cr} to
the case of constant curvature spaces.

The extension amounts to introducing new variables $Y^A$, momenta
$\bar P_B$ and postulating the following Poisson bracket relations on
the extended space:
\begin{equation}
  \pb{\bar P_A}{X^B}=-\delta_A^B\,,\qquad \pb{P_A}{Y^B}=-\delta_A^B\,.
\end{equation}
The original phase space can then be identified with the constrained
surface determined by the following second class constraints:
\begin{equation}
  P_A-\bar P_A=0\,,\qquad Y^A=0\,.
\end{equation}
Indeed, computing the Dirac bracket and solving constraints one
arrives at the original phase space.  Taking into account the original
constraints $L,M$ one finds that the equivalent set of constraints is
given by
\begin{equation}
\label{eq:const}
  P_A-\bar P_A=0\,,\qquad Y^A=0\,,\qquad  (X^A+Y^A)P_A=0\,, \qquad P^AP_A=0\,.
\end{equation}
One then observes that all constraints with $Y^A=0$ excluded are first class.

Passing to the quantum description one treats the variables $P,\bar P,
X, Y$ as quantum operators with the
following commutation relations\footnote{For later convenience, we
  deviate from the
  standard convention, used for instance in \cite{\BGST}, and choose
  ${\rm momenta}^{\rm here}= \imath\, {\rm momenta}^{\rm stand}$. We
  will also use $\brst^{\rm here}=\imath\brst^{\rm stand}$ below.}
\begin{equation}\label{eq:nonstand}
  \commut{\bar P_A}{X^B}=-\delta_A^B\,,\qquad
  \commut{P_A}{Y^B}=-\delta_A^B\,,\qquad
\end{equation}
and introduces the Grassmann odd ghost variables $\Theta^A$, $\mu$,
$c_0$ with $\gh{\Theta^A}=\gh{\mu}=\gh{c_0}=1$
and their conjugate momenta:
\begin{equation}
  \commut{b_0}{c_0}=-1\,,\qquad \commut{\rho}{\mu}=-1\,, \qquad
  \commut{\cP_A}{\Theta^B}=-\delta_A^B\,.
\end{equation}
Finally, the nilpotent BRST operator that takes into account the
first-class subset of~\eqref{eq:const}
reads as
\begin{equation}
\label{eq:brst-particle}
  \brst=\Theta^A( P_A- \bar P_A)+c_0 P^2+\mu(X^A+Y^A)P_A-2c_0\mu b_0\,.
\end{equation}

It can be worth mentioning that the easiest way to arrive
at~\eqref{eq:brst-particle} is to start from the BRST operator
\begin{equation}
  \brst_{stand}=\Theta^A \bar P_A +c_0 P^2+\mu X^AP_A-2c_0\mu b_0\,,
\end{equation}
which decomposes into independent pieces: the BRST operator
$\Theta^A\bar P_A$ eliminating pure gauge variables $\bar P, Y,
\Theta, \cP$ and the standard BRST operator for constraints $L,M$. At
this stage, the commutation relations are standard: $\commut{Y^A}{\bar
  P_B}=\delta^A_B$ and $\commut{X^A}{P_B}=\delta^A_B$. The BRST
operator \eqref{eq:brst-particle} with the commutation relations
\eqref{eq:nonstand} are then obtained by performing the change of
variables $X\to X+Y$ and $\bar P_A \to P_A-\bar P_A$. The advantage of
the more involved arguments leading to~\eqref{eq:brst-particle} is
that they can be naturally generalized to the case of a curved phase
space and to the case where the allowed functional spaces for $X$ and
$Y$ variables are different. In particular they remain valid if one
allows for formal power series in $Y$ variables in observables and
wave functions.

Before reducing to the surface by imposing the geometrical constraint
$X^2+l^2=0$, it is convenient to recast the system in a more
geometrical way.  This is achieved by introducing arbitrary
coordinates $\ul{X}^\ul{A}$ through $X^A=X^A(\ul{X}^\ul{B})$. An
$\ul{X}^\ul{B}$ dependent rotation in the fiber, $Y^{\prime}=\Lambda
Y$, $P^\prime= P{\Lambda^{-1}}$ can then be completed to a canonical
transformation if $\bar P_\ul{A}=\bar P^\prime_\ul{A}
-W^B_{\ul{A}C}Y^{\prime C} P^\prime_B$. After dropping the primes, the
transformed BRST operator becomes
\begin{equation}
\label{eq:brst-cov}
  \brst=\Theta^\ul{A}(E^B_{\ul{A}} P_B-\bar P_\ul{A})
+\Theta^\ul{A} W^C_{\ul{A}B}Y^BP_C+c_0P^2+
\mu(V^A+Y^A)P_A-2c_0\mu b_0\,,
\end{equation}
where
\begin{eqnarray}
  \label{eq:10}
W=-(d\Lambda) \Lambda^{-1}\,,\qquad V=\Lambda X\,,\qquad E=\Lambda
dX\,.
\end{eqnarray}

More generally, instead of $\mathbf R^{\dmn+1}$ one can consider a
$\dmn+1$ flat pseudo-Riemannian manifold $\manX$ with coordinates $
\ul{X}^{\ul{A}}$ and introduce $\bundle{\cV}(\manX)$, the vector
bundle associated with the orthonormal frame bundle and isomorphic to
$T(\manX)$. Fiber coordinates on $\bundle{\cV}(\manX)$ are denoted by
$Y^A$ and the flat fiber metric is $\eta=diag(-1,1,\dots,1,-1)$. The
fiber-wise isomorphism (vielbein) between $\bundle\cV(\manX)$ and
$T\manX$ is denoted by $E$.  Finally, one further extends the phase
space to the cotangent bundle $T^*(\bundle\cV(\manX))$, with
variables $\bar P_{\ul{A}},P_A$ being coordinates on the fibers.

At the quantum level the operators satisfy the commutation relations
\begin{equation}
  \commut{\bar P_{\ul{A}}}{\ul{X}^{\ul{B}}}=-\delta_{\ul{A}}^{\ul{B}}\,,\qquad
  \commut{P_A}{Y^B}=-\delta_A^B\,,
\end{equation}
originating from the canonical Poisson brackets on
$T^*(\bundle\cV(\manX))$.  The BRST operator \eqref{eq:brst-cov} is
nilpotent and (at least locally) describes a particle on a submanifold
$\manX_0\subset\manX$ transversal to the vector field $V^A
E_A^{\ul{B}}\dl{\ul{X}^\ul{B}}$ provided (i) the connection $W$ on
$\bundle{\cV}(\manX)$ is flat and compatible with the fiber-wise
metric and (ii) the vielbein is nondegenerate, covariantly constant,
and given by the covariant derivative of a fixed section $V$ of
$\bundle{\cV}(\manX)$:
\begin{equation}\label{eq:flat}
  \begin{aligned}
    W\eta+\eta W^T&=0\,,&\qquad  &dW+W^2=0\,,\\
    dE+WE&=0\,,&\qquad &E=dV+WV.
\end{aligned}
\end{equation}

The extension just described is a very simple example of so-called
Fedosov quantization~\cite{Fedosov-book} extended to the case of
constrained systems.  More precisely, it corresponds to a version of
Fedosov quantization adapted to the case of cotangent bundles which
was first considered in~\cite{Bordemann:1997er}. The extension to the
case of systems with constraints and the interpretation in terms of
BRST theory were developed in~\cite{Batalin:1989mb,\GL,\BGL}.  From
this perspective a natural generalization is to allow the connection
$W$ to be non-flat which can be appropriate when considering curved
phase space. In the case at hand taking $W$ flat is the simplest and
natural option. Curvature will now be introduced by restricting to a
submanifold.

\subsection{Reduction to the surface}

The essential feature of the extended system described by
\eqref{eq:brst-particle} is that the space-time coordinates $X^A$ and
their associated momenta are pure gauge degrees of freedom. It should
be noted, however, that the elimination of these degrees of freedom is
in general valid only locally and that geometrical data is lost in the
process. Moreover, at the level of associated field theories such an
elimination leads to theories which are not equivalent as local field
theories, as it relates for instance theories that live in different
space-time dimensions.  This is not so for the coordinate $r$
transversal to the surface $X^2+l^2=0$, which is not considered as a
proper space-time coordinate from the very beginning because by
construction the system effectively lives on this surface. In this
sense $r$ can be consistently eliminated both at the first-quantized
level and at the level of the associated field theory.

More technically, we first take a coordinate system
$\ul{X}^{\ul{A}}=(x^\mu,r)$ adapted to the surface: $r=\sqrt{-X^2}$ and
$X^A\dl{X^A}x^\mu=0$. In the new coordinate system,
the BRST operator takes the form:
\begin{multline}
  \brst=\theta^{(r)}(E_\r^AP_A-\bar p_{(r)})+\theta^\mu
  (E_{\mu}^AP_A-\bar
  p_\mu)+\theta^{(r)}W^A_{(r)\,B}Y^BP_A+\\+\theta^\mu
  W^A_{\mu\,B}Y^BP_A
+ c_0 P^2+\mu(V^A P_A+Y^AP_A)-2c_0\mu b_0\,,\label{eq:partcharge1}
\end{multline}
with $E,W,V$ defined as in \eqref{eq:10}, $V^2=-r^2$, and
  superscript $(r)$ denoting the component along $\dl{r}$ of a
tangent vector.

Suppose the system to be quantized in the coordinate representation
for the variables $r,\theta^{(r)},\bar p_{(r)},\cP_{(r)}$. In a
neighborhood of $r=l$, the variables $\theta^{(r)}$ and $r-l$ form
contractible pairs, or in other words, condition $r=l$ can be
considered as a gauge fixing condition.  It follows that the system
can be reduced by solving the linear second class constraints
$r=l,\theta^{(r)}=0,\cP_{(r)}=0,\bar p_{(r)}=0$ in the BRST operators
and putting $\theta^{(r)}=0$ and $r=l$ in the wave functions.  The
reduced BRST operator is given by
\begin{equation}
\label{eq:parent}
  \brst^\T=\theta^\mu(e^A_{\mu}P_A-\bar p_\mu)+\theta^\mu \omega^A_{\mu\,B}Y^BP_A
+c_0 P^2+\mu(V^A+Y^A)P_A-2c_0\mu b_0\,,
\end{equation}
where now $\omega^A_{\mu\,B}=W^A_{\mu\,B}(x,l)$,
$e^A_\mu=E^A_\mu(x,l)$, $V^A=V^A(x,l)$ and satisfy
\begin{equation}
\label{eq:EWV-comp}
\begin{aligned}
  d\omega_{\,A}^B+\omega^B_{\,C}\omega^{\,C}_A&=0\,,\qquad &
d e^A+\omega_{\,B}^A e^B&=0\,,\\
d V^A+\omega^A_{\, B} V^B&=e^A,\qquad &
 V^AV_A+l^2&=0 \,.
\end{aligned}
\end{equation}
In terms of the associated field theory, the fields associated to all
$r,\theta^{r}$ dependent states are then generalized auxiliary fields
in the sense of~\cite{\BGST}, provided one considers $r$ as an internal
degree of freedom rather than a space-time coordinate. A more detailed
proof of this fact can be found in~\cite{Barnich:2005ru} in the
nonlinear case\footnote{The parent formulation just described can be
  understood as a generalization of the unfolded
  formulation~\cite{Vasiliev:1988xc,Vasiliev:1988sa,Vasiliev:1994gr}
  in which the auxiliary role of space-time coordinates has been
  understood in~\cite{Vasiliev:2001zy}.}. Note that, contrary to the
other reductions considered in this paper, the one given here  
merely serves to define and motivate the parent theory on AdS and  
does not mean that the parent theory on AdS is equivalent, as a  
local field theory, to the flat theory in one dimension more.

The representation space $\Gamma(\bundle{\cH}^{\T})$ for the quantum
system is chosen to be the space of ``functions'' in
$x,Y,c_0,\mu,\theta$ which are formal power series in $Y$ with
coefficients in smooth functions in $x$ and polynomials in the
ghosts. In terms of the representation space 
the BRST operator acts as follows:
\begin{equation}
  \brst^\T=\derham-\theta^\mu \omega^A_{\mu\,B}Y^B\dl{Y^A}-\theta^\mu
  e_\mu^A \dl{Y^A} 
+c_0 {\ffrac{\partial^2}{\d Y^A\d Y_A}}-\mu(V^A+Y^A)\dl{Y^A}-2\mu
c_0\dl{c_0}, 
\end{equation}
where $\derham=\theta^\mu\dl{x^\mu}$ can be considered as the De Rham
differential provided one identifies $\theta^\mu$ and $dx^\mu$.  With
this choice, the quantum system described by the BRST
operator~\eqref{eq:parent} is a parent system for a scalar particle on
AdS. Indeed, we will now show that it can be reduced both to the
standard and the unfolded description of a particle on AdS.

To proceed with the reductions, we first note that a particularly
useful choice of $V$ is $V^{A}=l\delta^A_{(d)}$ for which
$e^{A}=l\omega^{A}_{(d)}$ implying
$e^{(d)}=0$.  Here and in what follows $Z^{(d)}$ denotes the $d+1$-th
component of a section $Z$, e.g.
$Z^AZ_A=Z^aZ_a+Z^{(d)}Z_{(d)}=Z^aZ_a-Z^{(d)}Z^{(d)}$.  With this
choice,
\begin{equation}
\nonumber
 \omega^A_{\,B}=  \left(\begin{array}{ll} \omega^a_{\,b} &
  \frac{1}{l} e^a \\  \frac{1}{l} e_b & 0
\end{array}\right)\,,
\end{equation}
and $e^{a}$, $a=0,\dots,d-1$, is the standard AdS vielbein, with
associated connection $\omega^a_{\,\,\,b}$:
\begin{equation}
  \begin{gathered}
  \label{eq:8}
  ds^2_{AdS}=\eta_{ab}\,e^a\otimes e^b,\qquad
  de^a+\omega^a_{\,\,\,b}e^b=0,\\
  R^{ab}\equiv d\omega^{ab}+
  \omega^a_{\,\,\,c}\omega^{cb}=-\frac{1}{l^2}e^ae^b\,,
\end{gathered}
\end{equation}
 and the BRST operator becomes
  \begin{equation}
  \label{eq:2}
  \brst^\T=\theta^\mu(e^a_{\mu}P_a-\bar p_\mu)+
\theta^\mu \omega^A_{\mu\,B}Y^BP_A
+c_0 P^2+\mu(lP_{(d)}+Y^AP_A)-2c_0\mu b_0.
\end{equation}

\subsection{Reduction to standard description}

It is straightforward to show that the system
$(\brst^\T,\Gamma(\bundle{\cH}^{\T}))$ can
be consistently reduced to the standard description of a particle on
AdS or, more precisely, to the system
$(\tilde\brst,\Gamma(\bundle{\cE}))$ where $\Gamma(\bundle{\cE})$ is
the space of $x,c_0$ dependent functions and
\begin{eqnarray}
  \label{eq:6}
  \tilde\brst{{\bundle\phi}_0}=
  c_0\Box_{AdS}{{\bundle\phi}_0},\qquad
  \Box_{AdS}{\bundle\phi}_0=\eta^{\mu\nu}(\d_\mu\d_\nu-\Gamma_{\mu\nu}^\rho\d_\rho)
  {\bundle\phi}_0
\end{eqnarray}
for ${{\bundle\phi}}_0\in\Gamma(\bundle{\cE})$ and where
$\Gamma^\rho_{\mu\nu}=e^\rho_a\omega_{\mu b}^a e_\nu^b+e^\rho_a\d_\mu
e^a_\nu $.
Details can be found in Appendix~\bref{sec:B1}.

\subsection{Reduction to unfolded form}

According to~\cite{\BGST}, the parent system can be reduced to its
unfolded form by computing the cohomology of the part of the BRST
differential that does not involve space-time ghosts
$\theta^\mu$.  One takes as degree minus the target space ghost
number, i.e., minus the degree in $c_0,\mu$, according to which the
BRST operator decomposes as $\brst= \brst_{-1}+\brst_0$, with
\begin{equation}
  \brst_{-1}=c_0
  \Box+\mu h-2\mu c_0\dl{c_0}, \qquad
  \brst_0=\derham-\omega^{A}_{\,\,B}Y^B\dl{Y^A}+\sigma\,,\label{eq:brstpartunf}
\end{equation}
where
  \begin{equation}
  \label{eq:5}
h=-(Y^A+V^A)\dl{Y^A}\,, \qquad \Box{} =\dl{Y^A}\dl{Y_A}\,,
\qquad \sigma =- \theta^\mu e_\mu^A \dl{Y^A}\,.
\end{equation}
For later convenience, let us set $Y^{(d)}=lz$ and $y^a=Y^a$. In
Appendix ~\bref{sec:B2}, we will prove:
\begin{prop}
The cohomology of $\brst_{-1}$ in the spacce $\cH^\T$ of formal power series
in $Y^A$ with coefficients in polynomials in $c_0,\mu,\theta^\nu$ is given by
\begin{equation}
  H^0(\brst_{-1},\cH^\T)\simeq \cE\subset \ker \brst_{-1}\,,\qquad
  H^n(\brst_{-1},\cH^\T)=0\quad n>0\,,
\end{equation}
where $\cE=\ker\Box \cap \ker h$. In the frame where
$V^A=l\delta^A_{(d)}$, $\cE$ is canonically isomorphic to the space
$\bar\cE$ of traceless $\mu,c_0,z$-independent elements. The
isomorphism $\lift\map\bar\cE\to\cE$ is given by
$\lift^{-1}\phi=\Pj(\phi|_{z=0})$ for any $\phi\in\cE$ where $\Pj$
denotes the projector to the traceless component (in the space of
$z$-independent elements if $\phi=\phi_0+(y^ay_a)\phi_1$ and
$\Box\phi_0=0$ then $\Pj\phi=\phi_0$).
\end{prop}
As recalled in Appendix~\bref{sec:A1}, the reduced BRST operators is
$\brst_0$ understood as acting in $\Gamma(\bundle\cE)$ because the
cohomology of $\brst_{-1}$ is concentrated in one degree. For ${\bundle\phi}\in
\Gamma(\bundle\cE)$ one then gets
\begin{equation}
\tilde\brst{\bundle\phi}=(\nabla+\sigma){\bundle\phi}\,, \qquad \nabla
=\derham-\omega^A_{\,\,B}Y^B\dl{Y^A}\,. 
\end{equation}
The associated equations of motion take the form
$(\nabla+\sigma)\Psi^{(0)}=0$ where $\Psi^{(0)}$ contains the fields
associated with the $\theta^\mu$-independent states in $\cE$, i.e.,
$\Psi^{(0)}=\Psi^{(0)}(x,Y)$ and $\Box\Psi^{(0)}=h\Psi^{(0)}=0$.

It is natural to rewrite the system in terms of $\bar\cE$ valued
fields.  More precisely, in the frame where $V^A=l\delta^A_{(d)}$ the
operators $h$ and $\brst_0$ take the form
\begin{equation}
\label{eq:h-brst-exp}
  h=-Y^a\dl{Y^a}-(z+1)\dl{z}\,, \quad \brst_0=\derham-
\omega^{a}_{\,\,b}Y^b\dl{Y^a}
+(z+1)\sigma-\frac{1}{l^2}e^a y_a\dl{z}\,.
\end{equation}
For any $\phi_0\in\bar\cE$, element $\phi=\lift\phi_0$ reads as (see
Appendix~\bref{sec:B2})
\begin{equation}
\label{eq:phi-exp}
  \phi=\frac{1}{(1+z)^n}(\phi_0+
  (y^ay_a)\frac{n(n+1)}{2l^2(d+2n)}\phi_0+\ldots)\,,
\end{equation}
where $n=y^a\dl{y^a}$, the ratio is understood as a formal power series, and
$\ldots$ denote terms of the form $(y^ay_a)^k \phi_0, \,\, k\geq 2$.

In terms of $\bar\cE$-valued sections, the reduced BRST operator is
given by $\brst^{\rm unf}=\lift^{-1}\tilde\brst\lift$.
Using~\eqref{eq:h-brst-exp} and \eqref{eq:phi-exp}, one finds as
explicit expression
\begin{equation}
\brst^{\rm unf} {\bundle\phi}_0 =[\derham -\omega^a_{\,\,b}y^b\dl{y^a}
+\sigma]{\bundle\phi}_0
-\frac{(n-1)(n+d-2)}{l^2(d+2N-2)}\Pj[e^ay_a {\bundle\phi}_0] \,,
\end{equation}
where ${\bundle\phi}_0\in \Gamma(\bundle{\bar\cE})$ and $\Pj$ denotes
the projector to the subspace of traceless elements. This expression
coincides with the differential determining the unfolded form of the
Klein-Gordon equation on AdS space proposed
in~\cite{Shaynkman:2000ts}.

\section{Free higher spin gauge fields on AdS}\label{sec:fronsdal}
\subsection{The first-quantized model}

Instead of the standard string inspired first-quantized description of
higher spin gauge fields in flat
\cite{Ouvry:1986dv,Bengtsson:1986ys,Henneaux:1987cp} and in AdS space
in intrinsic coordinates \cite{Buchbinder:2001bs,Sagnotti:2003qa}, we
follow here the strategy of the preceding section and construct a
parent theory for higher spin gauge fields on constant curvature
spaces by using an embedding. Namely, we incorporate the constraints
describing the reduction to the surface into the flat first-quantized
BRST system~\cite{Ouvry:1986dv,Bengtsson:1986ys,Henneaux:1987cp} in
the embedding space $\fR^{\dmn+1}$.

With respect to the particle, the additional variables besides
$X^A,P_B$ are
$a^A,a^{\dagger B}$, where $A=0,\dots,\dmn$. At the quantum level,
these variables satisfy the commutation relations
\begin{equation}
  \commut{P_B}{X^A}=-\delta^A_B,\qquad
  \commut{a^A}{a^{\dagger B}}=\eta^{AB}.
\end{equation}
The constraints of the system are
\begin{equation}
  \begin{aligned}
\label{eq:constraints}
\cL&\equiv\eta^{AB}P_A P_B=0,\qquad& T&\equiv\eta_{AB}a^A a^B=0,\\
\cS^\dagger&\equiv -P_A a^{\dagger A}=0,\qquad& \cS&\equiv -P_A a^A=0 .
\end{aligned}
\end{equation}
For the ghost pairs $(c_0,b_0)$, $(c^\dagger,b)$,
$(c,b^\dagger)$, and $\xi,\pi$ corresponding to each of these
constraints, we take
the canonical commutation relations in the form\footnote{We use the
  ``super'' convention that $(ab)^\dagger=(-1)^{\p{a}\p{b}}b^\dagger
  a^\dagger$.}
\begin{equation}
  \commut{b_0}{c_0}=-1, \qquad \commut{c}{b^\dagger}=1,\qquad
  \commut{b}{c^\dagger}=-1, \qquad \commut{\pi}{\xi}=-1\,.
\end{equation}
The ghost-number assignments are
\begin{equation}\label{target-gh-n}
  \begin{gathered}
        \gh{c_0}=\gh{c}=\gh{c^\dagger}=\gh{\xi}=1,\\
    \gh{b_0}=\gh{b}=\gh{b^\dagger}=\gh{\pi}=-1.
\end{gathered}
\end{equation}
The BRST operator is then given by
\begin{equation} \label{eq:Fbrst}
  \brst_0=c_0\cL+c^\dagger\cS+\cS^\dagger c+\xi T
 +c^\dagger c b_0+2\xi cb,
\end{equation}
while the representation space consists of functions in $X^A$ (on which
$P_A$ acts as $-\dl{X^A}$) with values in the ``internal space''
$\cH_0$.  The latter is the tensor product of the space
$\cH_{c_0,\xi}$ of functions in $c_0,\xi$ (coordinate representation
for $(c_0,b_0)$ and $(\xi,\pi)$) and the Fock space for $(a^\dagger_A,a^A)$,
$(c^\dagger,b)$, and $(c,b^\dagger)$ defined by
\begin{equation}\label{target-vac}
  a^A\vac=b\vac=c\vac=0.
\end{equation}

To reduce the system to AdS space, in addition to constraints
\eqref{eq:constraints}, one needs to impose the ``geometrical''
constraints
\begin{equation}
\label{eq:2-constraints}
X^2+l^2=0\,,\quad X^A P_A=0\,, \quad X^Aa_A=0\,, \quad X^Aa^\dagger_A=0\,,
\end{equation}
which are second class.

Again, after first passing to an equivalent set of second class
constraints, we will keep only half of them so that the remaining
constraints together with~\eqref{eq:constraints} form a first class
set, the other constraints being considered as partial gauge fixing
conditions. More precisely, one first considers the constraints
\begin{equation}
  X^AP_A+a^\dagger_A a^A=0\,,\qquad X^Aa_A=0
\end{equation}
and checks that together with constraints~\eqref{eq:constraints}, they
form a closed algebra.
Introducing new pairs of ghost variables $\mu,\rho$ and $\nu, \tau$ with
\begin{equation}
  \gh{\mu}=\gh{\nu}=1\,,\qquad \gh{\rho}=\gh{\tau}=-1
\end{equation}
and commutation relations
\begin{equation}
 \commut{\rho}{\mu}=-1 \,,\quad \commut{\tau}{\nu}=-1 \,,
\end{equation}
one can incorporate all the constraints into a standard BRST
operator. The resulting BRST system
\cite{Bengtsson:1990un,Bonelli:2003zu} describes Fronsdal's higher spin
gauge fields on AdS.

In order to construct the parent theory we introduce, as in the
previous section, new variables $Y^A$, momenta $\bar P_A$ and ghost
pairs $\Theta^A,\cP_A$ with commutation relations:
\begin{equation}
  \commut{\bar P_A}{X^B}=-\delta_A^B\,,\qquad
  \commut{P_A}{Y^B}=-\delta_A^B\,,\qquad
  \commut{\cP_A}{\Theta^B}=-\delta_A^B\,.
\end{equation}
The extended system is now described by the constraints
\begin{equation}
\label{eq:p-constraints}
 P_A-\bar P_A=0\,,
\end{equation}
and all original constraints understood as functions of $P$ and
extended by $Y$-dependent terms so as to commute with
constraints~\eqref{eq:p-constraints}. In fact only constraints
$XP+a^\dagger a=0$ and $XA=0$ get corrected to $XP+XY+a^\dagger a=0$
and $Xa+Ya=0$ respectively. Finally, one constructs the
following nilpotent BRST operator:
\begin{multline}
  \brst=\theta^A(P_A -\bar P_A)+c_0P^2-a^\dagger P c-c^\dagger Pa+\xi
  a^2+\\+\mu[(X+Y)P+a^\dagger a]
+\nu(X+Y)a+\text{terms cubic in ghosts}\,.
\end{multline}

\subsection{Reduction to the surface and algebraic structure}

Proceeding exactly in the way as in the case of the scalar particle,
the BRST operator for the system pulled back to $\manX_0$ is
\begin{multline}
\label{eq:intrinsic}
  \brst^{\rm T}=\theta^{\mu}(e_{\mu}^B P_B- \bar p_{\mu})
+\theta^{\mu}\omega^B_{\mu\,C} (Y^C P_B-a^{\dagger C} a_B)
+\\+
c_0 \Box+\sd c+c^\dagger S+ \xi T
+
\mu h +\nu \bsd
+
\text{terms cubic in   ghosts}\,,
\end{multline}
where the following notations have been introduced for constraints
\begin{equation}
\label{eq:operators}
    \begin{gathered}
     S=-aP\,,\qquad S^\dagger=-a^\dagger P \,,\qquad T=a^2\,,\qquad \Box=P^2\,,\\
\bar S^\dagger=(Y+V)a\,,\qquad  h=a^\dagger a+(Y+V)P\,,
\end{gathered}
\end{equation}
and $e^A$, $V^A$, $\omega_A^B$ satisfy~\eqref{eq:EWV-comp}. The
constraint algebra that determines the form of the terms cubic in
ghosts reads as:
\begin{equation}
  \begin{gathered}\label{eq:algebra}
    \commut{S}{S^\dagger}=\Box\,,\quad  \commut{h}{\Box}=2\Box\,,\quad
    \commut{h}{S^\dagger}=2S^\dagger\,, \quad
  \commut{S}{\bar S^\dagger}=T\,,\quad
  \commut{S^\dagger}{\bar S^\dagger}=h\,,\\
\commut{h}{T}=-2T\,,\quad \commut{h}{\bar S^\dagger}=-2\bar S^\dagger\,,\quad
    \commut{T}{S^\dagger}=2S\,,\qquad
\commut{\Box}{\bar S^\dagger}=2S\,,
\end{gathered}
\end{equation}
with all other commutators vanishing. It is not difficult to see that
it is a subalgebra of $sp(4)$ identified in \cite{\BGST} as the
algebraic structure underlying the parent theory of Fronsdal higher
spin gauge fields in the flat space. Note however that in the flat
case the subalgebra of $sp(4)$ entering the BRST operator does not
contain the generators $\bar S^\dagger,h$ and all the $sp(4)$
generators are represented on variables $Y^\mu,a^{\dagger\mu}$
associated with $\dmn$-dimensional tangent space. In the case at hand
the variables $Y^A,a^{\dagger A}$ are associated with the $\dmn+1$
dimensional embedding space.

Another important difference with the flat case is the shift $Y\to
Y+V$ in the generators. Moreover, it follows from
$d V^A+\omega^A_{\, B} V^B=e^A$ that in terms of $Y^\prime=Y+V$ the
expression for the BRST operator takes the form
\begin{multline}
  \brst^{\rm T}=-\theta^{\mu}\bar p_{\mu}
+\theta^\mu\omega^B_{\mu\,C}
({Y^\prime}^C P_B-a^{\dagger C} a_B)
+
\\
+c_0 \Box+\sd c+c^\dagger S+
  \xi T+\mu h +\nu \bsd+
\text{terms cubic in   ghosts}\,.
\end{multline}
Because the BRST operator is polynomial in $Y$ the change of
coordinates $Y^\prime=Y+V$ is legitimate and gives, in particular, the
easiest way to check nilpotency explicitly. Indeed, in these terms,
nilpotency immediately follows from the fact that $W$ is a flat
connection compatible with the metric $\eta_{AB}$ and all generators
\eqref{eq:operators} are build from $a^\dagger,a$ and $Y^\prime,P$
with the indexes contracted with $\eta_{AB}$. Note that in general one
is not allowed to do such a change of variables in the representation
space where $Y,P$ are represented on formal power series in $Y$.

In order to obtain a representation of all of $sp(4)$, one needs to
add the following generators:
\begin{equation}
\bar T=-\frac{1}{4}(a^\dagger)^2\,,\quad \bar S=(Y+V)a^\dagger
\,,\quad\bar \Box=(V 
+Y)^2\,,\quad h^\prime=-a^\dagger a-\frac{\dmn+1}{2}\,,
\end{equation}
which makes the total number of generators ten as it should be.

Taking as representation space $\Gamma(\bundle{\cH}^{\T})$, the space
of ``functions'' in variables $x,Y,a^\dagger$,
$c^\dagger,c_0,b^\dagger$, $\xi,\mu,\nu,\theta^\mu$ which are formal
power series in the variables $Y^A$, smooth functions in $x$, and
polynomials in $a^{\dagger A}$ and ghost variables, the quantum system
described by $\brst^\T$ is the parent system for free higher spin
gauge fields on AdS. The action of the BRST
operator~\eqref{eq:intrinsic} on a generic state of
$\phi\in\Gamma(\bundle{\cH}^{\T})$ takes the form
\begin{equation}
\label{eq:brst-T}
  \brst^{\T}\phi=(\nabla+\sigma+\bar\brst)\phi\,,
\end{equation}
where
\begin{equation}
   \nabla=\derham-\omega_{\,\,
     A}^BY^A\dl{Y^B}-\omega_{\,\, A}^B
a^{\dagger A}\dl{a^{\dagger B}}\,,\qquad \sigma =-  \theta^\mu e_\mu^A \dl{Y^A}\,,
\end{equation}
and
\begin{multline}
\label{eq:bar-brst}
 \bar \brst=
c_0 \Box
+ c^\dagger S
+ S^\dagger \dl{b^\dagger}
+\xi T
+
\mu (h-2)
+
\nu \bar S^\dagger
-c^\dagger \dl{b^\dagger}\dl{c_0}
-2\xi\dl{b^\dagger}\dl{c^\dagger}
-
\\
-2\mu c_0\dl{c_0}
+2\mu b^\dagger\dl{b^\dagger}
+2\mu\xi\dl{\xi}
+2\nu c_0\dl{c^\dagger}
+2\mu \nu \dl{\nu}
+\nu c^\dagger\dl{\xi}
+\nu\dl{b^\dagger}\dl{\mu}\,.
\end{multline}
The operators $\Box,T,S,\bsd,\sd,h$ are given by \eqref{eq:operators}
with $P_A,a_A$ replaced with $-\dl{Y^A}$ and $\dl{a^{\dagger A}}$
respectively.  By various consistent reductions of the parent system
$(\brst^\T,\Gamma(\bundle\cH^\T))$, one can reach the original and the
unfolded descriptions as well as some ``intermediate'' formulations
which can be interesting in their own right.

\subsection{Reduction to standard description}\label{sec:4.3}

As a first step, a new intermediate reduction could turn out to be
useful. It preserves the simple algebraic structure of the parent
theory and involves the $d+1$ oscillators while eliminating all the
$Y$-variables but one.

\subsubsection{Tensor fields in embedding space}
\label{sec:tens-fields-embedd}

The reduction consists in the elimination of $Y^a,\theta^\mu$. In this
case, $\Gamma(\bundle{\cE})$ is the space of functions in $x$ with
values in formal power series in $Y^{(d)}$ and polynomials in
$a^{\dagger A}$, $c_0$, $c^\dagger$, $b^\dagger$, $\mu$, $\nu$.
We again choose $V^A=l\delta^A_{(d)}$ and set $Y^{(d)}=lz$,
$a^{\dagger (d)}=lw$.  Let
$\partial_a=e^\mu_a\frac{\partial}{\partial x^\mu}$,
$\omega^B_{a\,C}=e^\mu_a\omega^B_{\mu\,C}$ and
\begin{equation}\label{eq:20}
  \cD_a=\partial_a-\omega^B_{a\,C}a^{\dagger C}\dl{a^{\dagger B}}
=\d_a-
\omega^b_{a\,c}a^{\dagger c}\dl{a^{\dagger b}}
-w\dl{a^{\dagger a}}-
\frac{a^\dagger_a}{l^2}\dl{w}\,,
\end{equation}
so that $[\cD_a,\cD_b]=(\omega^c_{ab}-\omega^c_{ba})\cD_c$.  We show
in Appendix~\bref{sec:C1} that the reduced BRST differential is given
by
  \begin{multline}
  \label{eq:12}
  \tilde\brst=c_0\tilde\Box+\tilde
  S^\dagger\dl{b^\dagger}+c^\dagger\tilde S +\xi \tilde T +\mu \tilde
  h +\nu \tilde{\bar S}^\dagger
  -c^\dagger\dl{b^\dagger}\dl{c_0} -2\mu c_0\dl{c_0} - \\
  -2\mu \dl{b^\dagger} b^\dagger
  +2\mu\xi\dl{\xi}+2\mu\nu\dl{\nu}-2\xi\dl{b^\dagger}\dl{c^\dagger}
  +2\nu c_0\dl{c^\dagger}+
  \nu c^\dagger\dl{\xi}+\nu\dl{b^\dagger}\dl{\mu}\,,
\end{multline}
where
\begin{align}
  \label{eq:13}
  \tilde\Box &~=~\big(\frac{1}{1+z}\big)^2
 \eta^{ac}(\delta^b_c\cD_a-\omega^b_{a\, c})\cD_b
-\frac{1}{l^2}\dl{z}\dl{z}-\frac{\dmn}{l^2}
\frac{1}{1+z}
\dl{z}
\,, \\
\tilde S^\dagger&~=~\frac{1}{1+z}a^{\dagger a}
\cD_a+ w\dl{z}\,,\\
\tilde S &~=~\frac{1}{1+z}
\dl{a^{\dagger}_a}\cD_a-\frac{1}{l^2}\dl{w}\dl{z}\,,\\
\tilde h &~=~a^{\dagger A}\dl{a^{\dagger A}}-(1+z)\dl{z}\,,\\
{\tilde{\bar S}}^\dagger &~=~(1+z)\dl{w}\,,\\
\tilde T &~=~\dl{a^\dagger_A}\dl{a^{\dagger A}}\,,
\end{align}
and the operators $\tilde\Box$, $\tilde S$, $\tilde S^\dagger$,
$\tilde T$, $\tilde h$, ${\tilde{\bar S}}^\dagger$ satisfy the same subalgebra
\eqref{eq:algebra} of $sp(4)$ as the corresponding untiled
operators.

One can reduce further by eliminating the dependence on $\xi$. This
can be done consistently by restricting the remaining states,
respectively the string fields, to be annihilated by
$\tilde\cT_0=\dl{a^\dagger_A}\dl{a^{\dagger
    A}}-2\dl{b^\dagger}\dl{c^\dagger}$ and dropping all terms in the
BRST operator that involve $\xi$ or $\dl{\xi}$. Indeed, by choosing as
a degree minus the homogeneity in $\xi$, the lowest part of the BRST
operator \eqref{eq:12} is $\tilde\brst_{-1}=\xi \tilde\cT_0$. Its
cohomology is concentrated in degree $0$, the $\xi$ independent part,
and described by states annihilated by $\tilde\cT_0$. It then follows
directly from Proposition~\bref{prop:red} that the reduced BRST
differential is given by $\tilde\brst_{\xi=0=\pi}$ restricted to the
subspace of $\xi$-independent and $\tilde\cT_0$-traceless elements.
Here, $\tilde\brst_{\xi=0=\pi}$ denotes the BRST operator
$\tilde\brst$ with all $\xi,\dl{\xi}$-dependent terms dropped.

Note that one can consider the BRST operator $\tilde\brst_{\xi=0=\pi}$
as an operator acting in the subspace of $\xi$-independent elements
that are not necessarily annihilated by $\tilde\cT_0$. However, this
operator is not strictly nilpotent anymore, nor does it commute with
$\tilde\cT_0$. More precisely, it satisfies
$[\tilde\cT_0,\tilde\brst_{\xi=0=\pi}]=O\tilde\cT_0$,
$\tilde\brst_{\xi=0=\pi}^2=P\tilde\cT_0$ for some operators $O,P$, as
it should for $\tilde\brst_{\xi=0=\pi}$ to be well-defined and
nilpotent on the $\tilde\cT_0$-traceless subspace.

Contrary to the case of higher spin gauge fields in flat space, one
can thus not directly remove the trace constraint at the level of the
parent theory or the intermediate reduction by simply imposing it on
the states and the string fields and dropping $\xi,\dl{\xi}$ dependent
terms in the BRST operator. The reason is that the commutator
$\commut{S}{\bsd}=T$ of operators $S$ and $\bsd$ entering the BRST
operator produces $T$, and similarly for the tiled operators.

\subsubsection{Tensor fields on AdS}
\label{sec:tensor-fields-ads}

We now consider the reduction to the standard BRST description in
intrinsic coordinates, i.e., with $z,w,\mu,\nu$ eliminated. The system
is described by the space $\Gamma(\bundle{\cE})$ of functions in
$x^\mu$ taking values in polynomials in $a^{\dagger
  a},c_0,c^\dagger,b^\dagger,\xi$. In this case, we show in the
appendix that the BRST operator reduces to
  \begin{equation}
  \label{eq:16}
 \hat\brst=c_0\hat\Box+c^\dagger \hat S+\hat S^\dagger\dl{b^\dagger}
-c^\dagger\dl{b^\dagger}\dl{c_0}+\xi \hat \cT\,,
\end{equation}
where
\begin{equation}
  \begin{aligned}
  \hat\Box &~=\eta^{ac}
(\delta^b_cD_a-\omega^b_{a\, c})D_b + \frac{1}{l^2}
\big(N_{a^{\dagger^a}} +2c^\dagger\dl{\xi} a^{\dagger a}\nabla_a+
\\
 &\quad+(3-\dmn-N_{a^{\dagger a}}-2N_{b^\dagger}-2N_\xi)
(N_{a^{\dagger a}}-2+2
N_{b^\dagger}+2N_{\xi}) \big)
\,,
 \\
\hat S &~=\dl{a^{\dagger}_a}D_a+\frac{1}{l^2}
\big(2c_0\dl{c^\dagger}(2N_{a^{\dagger a}}+\dmn-3+2N_{b^\dagger} +2
N_\xi)\big)\,,
\\
\hat S^\dagger&~=a^{\dagger a}
D_a+\frac{1}{l^2} a^{\dagger
a}a^{\dagger}_a(c^\dagger\dl{\xi}+2c_0\dl{c^\dagger})\,,
\\
\hat \cT &~=\dl{a^\dagger_a}\dl{a^{\dagger
a}}-2\dl{b^\dagger}\dl{c^\dagger}-\frac{1}{l^2}2c_0(1-2N_{c^\dagger})\dl{\xi}\,,
\end{aligned}
\end{equation}
with $N_{Z^i}=Z^i\dl{Z^i}$ for any variables $Z^i$ and
\begin{eqnarray}
D_a=\partial_a-\omega^b_{a\,c}a^{\dagger c}\dl{a^{\dagger b}},\,
  [D_a,D_b]=(\omega^c_{ab}-\omega^c_{ba})D_c
-\frac{1}{l^2}(
  a^{\dagger}_a\dl{a^{\dagger b}}- a^{\dagger}_b\dl{a^{\dagger
  a}})\,.
\end{eqnarray}

Finally, one can reduce further by eliminating the dependence on
$\xi$. Exactly the same reasoning as for the analogous reduction in
the previous subsection shows that this can be done consistently by
restricting the remaining states, respectively the string fields, to
be annihilated by $\hat\cT_0=\dl{a^\dagger_a}\dl{a^{\dagger
    a}}-2\dl{b^\dagger}\dl{c^\dagger}$ and dropping all terms in the
BRST operator that involve $\xi$ or $\dl{\xi}$. We denote by
$\hat\brst_{\xi=0=\pi}$ the BRST operators $\hat\brst$ with all the
$\xi,\dl{\xi}$ dependent terms dropped. 

As before, one can consider $\hat\brst_{\xi=0=\pi}$ as an extension of
the reduced BRST operator from the subspace of $\xi$-independent and
$\hat\cT_0$-traceless elements to that of $\xi$-independent, but not
necessarily $\hat\cT_0$-traceless ones.  A more convenient extension,
however, turns out to be
\begin{equation}
\brst_{\rm mod}=\hat\brst_{\xi=0=\pi}
+\frac{c_0}{l^2}(a^{\dagger  a}a^\dagger_a+4c^\dagger b^\dagger)\hat \cT_0\,,
\end{equation}
because $\brst_{\rm mod}$ is strictly nilpotent and commutes with
$\hat\cT_0$.

Up to conventions and an overall sign, $\brst_{\rm mod}$ coincides
with the hermitian BRST operator constructed in this context in
\cite{Buchbinder:2001bs} (see also \cite{Sagnotti:2003qa}).  By
constraining the string field to be annihilated by $\hat \cT_0$ and
assuming the appropriate reality condition, the master action for
higher spin fields on AdS is given by (\ref{eq:psiopsi}), using either
$\brst_{\rm mod}$ or $\hat\brst_{\xi=0=\pi}$. More details on master
actions of this type can be found for instance in
\cite{Barnich:2003wj,\BGST,Barnich:2005bn}.

It then follows from Subsection~\bref{sec:Lagrangian} that all the
formulations of higher spin gauge fields on AdS described in this
paper are Lagrangian by using suitable generalized auxiliary fields.

\subsection{Reduction to unfolded form}

The reduction to the unfolded form is performed by reducing to the
cohomology of $\bar\brst$ given by~\eqref{eq:bar-brst}. As in the flat
case~\cite{\BGST}, we do this reduction in several steps.

\subsubsection{Reducing to totally traceless fields}

First we reduce the parent theory to a theory with totally traceless
fields.  This is achieved by taking as a degree minus the homogeneity
in $c_0,c^\dagger,\xi$ and reducing to the cohomology of the part of
$\bar\brst$ in lowest degree $-1$,
\begin{equation}
\label{eq:brst-trace}
  \brst_{\rm trace}=c_0 \Box+ c^\dagger S+ \xi T\,.
\end{equation}
The dimension of the space in which the trace is taken is $d+1$, the
dimension of the embedding space. In $d+1\geq 3$, the cohomology of
$\brst_{\rm trace}$ in ${\cH}^\T$, the space of formal power series in
variables $Y^A$ with coefficients in polynomials in $a^{\dagger A}$
and ghost variables, is given by~\cite{\BGST}:
\begin{equation}
  \begin{aligned}
    H^0(\brst_{\rm trace},\cH^\T)&\cong\tilde\cE=\{\phi\in \cH^\T \map
    \Box\phi=S\phi=T\phi=0,\,\,{\rm deg} (\phi)=0\}\,,\\
 H^n(\brst_{\rm trace},\cH^\T)&=0\qquad n
    \neq 0\,.
\end{aligned}
\end{equation}
Since the cohomology is concentrated in one degree, one immediately
arrives at:
\begin{prop}
  The parent system $(\brst^\T,\Gamma(\bundle{\cH^\T}))$ can be
  consistently reduced to the system
  $(\tilde\brst^\T,\Gamma(\bundle{\tilde\cE}))$ with
  $\tilde\brst^\T=\nabla+\sigma+\tilde\brst$ and
\begin{equation}
\label{eq:sl2brst}
\tilde\brst~=~  S^\dagger\dl{b^\dagger}+\mu(h-2) +\nu \bar
S^\dagger+2\mu b^\dagger \dl{b^\dagger}
+2\mu\nu \dl{\nu}+\nu \dl{b^\dagger}\dl{\mu}\,.
\end{equation}
\end{prop}
The associated field theory is described by the physical fields
$F,H,G,A$ taking values in $Y,a^\dagger$-dependent traceless elements
and entering the ghost number zero component of the string field
\begin{equation}
  \tilde\Psi^{(0)}=F+\mu b^\dagger H+\nu
  b^\dagger G
  +b^\dagger \theta^\mu A_\mu\,.
\end{equation}
In terms of component fields, the equations of motion
$\tilde\brst^\T\tilde\Psi^{(0)}=0$
read as
\begin{equation}
  \begin{gathered}
    DA=0, \qquad D F+\sd A=0,\qquad D H+h A=0,\qquad DG+\bsd A=0,\\
    (h-2)F-\sd H=0,\qquad \bsd F+H-\sd G=0,\qquad (h+2)G-\bsd H=0\,,
\end{gathered}
\end{equation}
where $D=\nabla+\sigma$. The gauge symmetries are determined by
$\delta \tilde\Psi^{(0)}=\tilde\brst^\T\tilde\Psi^{(1)}$ with component
fields in $\tilde\Psi^{(1)}$ to be replaced with gauge parameters.

The next step is to reduce to the cohomology of the BRST
operator~\eqref{eq:sl2brst}.  This operator corresponds to the
standard Chevalley-Eilenberg differential associated with the Lie
algebra $sl(2)$ in the given representation.

If $V^A$ were vanishing, the Lie algebra would act homogeneously in
$Y,a^\dagger$ and the representation space would split into the direct
sum of finite-dimensional irreducible representations. In this case,
the cohomology is well known and given by the Lie algebra invariants
in ghost numbers $-1,1$. In our case, however, the operators act
inhomogeneously so that infinite-dimensional representations have to
be taken into account.  In particular, there can be nontrivial
cohomology classes associated with elements which are not polynomial
but are formal power series in $Y$. It is again instructive to split
the reduction to the cohomology of $\tilde\brst$ into two steps.

\subsubsection{First step: reduction to the intermediate system}
Taking as a degree minus the homogeneity in ghost variables $\mu,\nu$,
one arrives at the decomposition
$\tilde\brst=\tilde\brst_{-1}+\tilde\brst_0$,
where
\begin{equation}
\tilde\brst_{-1}=\mu(h-2) +\nu \bar S^\dagger+2\mu b^\dagger \dl{b^\dagger}
+2\mu\nu \dl{\nu}\,,\qquad \tilde\brst_0=S^\dagger\dl{b^\dagger}+\nu
\dl{b^\dagger}\dl{\mu}\,.
\end{equation}
In order to carry out the first step of the reduction, we need:
\begin{prop}\label{prop:4inter}
The cohomology of $\tilde\brst_{-1}$ in $\tilde\cE$ is given by
\begin{equation}
  H^0(\tilde\brst_{-1},\tilde\cE)=\hat\cE\,,\qquad
  H^n(\tilde\brst_{-1},\tilde\cE)=0\quad n\neq 0,
\end{equation}
where $\hat\cE\subset\tilde\cE$ is the subspace of
$\mu,\nu$-independent elements satisfying
\begin{equation}\label{eq:gn0-cond}
(h-2+2b^\dagger\dl{b^\dagger})\phi=0\,, \qquad
\bar S^\dagger \phi=0\,.
\end{equation}
\end{prop}
In degree zero, the statement is trivial. That the cohomology of
$\tilde\brst_{-1}$ vanishes in nonzero degree is shown in
Appendix~\bref{sec:C2}. Because the cohomology of $\tilde\brst_{-1}$
is concentrated in one degree, the reduction is straightforward:
\begin{prop}
  The system $(\tilde\brst^\T,\Gamma(\bundle{\tilde\cE}))$ can be
  consistently reduced to the \textit{intermediate system}
  $(\hat\brst^\T,\Gamma(\bundle{\hat\cE}))$ where $\hat\cE$ is described
  by~\eqref{eq:gn0-cond} and
\begin{equation}
\label{eq:hat-brst}
  \hat\brst^\T=\nabla+\sigma+\sd\dl{b^\dagger}\,.
\end{equation}
\end{prop}
Note that $\nabla+\sigma$ commutes with $\bsd$ and
$h-2+2b^\dagger\dl{b^\dagger}$.
Similarly, $\sd\dl{b^\dagger}$ preserves $\hat\cE$ and therefore
projectors are not needed in~\eqref{eq:hat-brst}.

The equations of motion of the associated free field theory have the form
\begin{equation}
  (\nabla  +\sigma) \hat A=0\,,\qquad (\nabla +\sigma) \hat F=-\sd \hat A\,.\label{eq:int}
\end{equation}
Here $\hat A=\theta^\mu \hat A_\mu(x;Y,a^\dagger)$ and $\hat
F(x;Y,a^\dagger)$ are respectively the 1-form and the 0-form physical
fields entering the string field associated with $\hat\cE$:
\begin{equation}
  \hat\Psi^{(0)}=\hat F+b^\dagger\theta^\mu \hat A_\mu\,.
\end{equation}
The gauge transformations are given by
\begin{equation}
  \delta \hat A=(\nabla + \sigma)\hat \lambda\,,\qquad
 \delta \hat F=-\sd\hat\lambda\,,
\end{equation}
where the gauge parameter $\hat\lambda(x;Y,a^\dagger)$ replaces the
fields at ghost number $1$ and therefore satisfies
$h\hat\lambda=\bsd\hat\lambda=0$.

This representation of higher spin gauge fields generalizes the
so-called \textit{intermediate form} identified in \cite{\BGST} to the
case of a constant curvature background.

\subsubsection{Second step: reduction to the unfolded form}

In the next step, we take as a degree the homogeneity in $b^\dagger$ so
that $\hat\brst_{-1}=\sd\dl{b^\dagger}$. To reduce the system, we need
to compute the cohomology of $\hat\brst_{-1}$ in the space $\hat\cE$
of completely traceless formal power series in $Y$ with coefficients
in polynomials in $a^\dagger,b^\dagger,\theta^\mu$
satisfying~\eqref{eq:gn0-cond}.

In degree $1$, the coboundary condition is trivial
while the cocycle condition gives $\sd \phi=0$.  This condition is
homogeneous in variables $Y,a^\dagger$ and implies that a homogeneous
component of $\phi$ has more $a^\dagger$ than $Y$
variables. Together with the condition $h\phi=0$, this
implies that $\phi$ is polynomial in $Y$ as well since otherwise the
condition $h\phi=0$ can be satisfied only by a formal
power series in $z$, which contradicts $\sd\phi=0$. Therefore any
solution can be decomposed into homogeneous solutions in $a^{\dagger
  A}$ and $Y^A+V^A$. These solutions are described by
traceless rectangular Young tableaux.

In degree $0$, the cocycle condition is trivial while the coboundary
tells us that $\phi \sim \phi+\sd A$ for a traceless $A$ satisfying
$hA=\bsd A=0$. In fact in each equivalence class there exists a unique
representative satisfying $V^A\dl{a^{\dagger A}}\phi=0$. Because the
statement does not depend on the choice of local frame it is enough to
show this in the frame where $V^A=l\delta^A_{(d)}$. As usually we use
notations: $lz=lz^1=Y^{(d)}$, $lw=lz^2=a^{\dagger (d)}$, and
$y^a=Y^a$. We need the following:
\begin{prop}\label{prop:hbsd-lift-2}
  For any $z^\alpha$ independent $\phi_0\in\tilde\cE$ and an
  arbitrary number $m$, there exists a unique solution $\phi\in\tilde\cE$
  satisfying the equation:
\begin{equation}
\label{eq:K-def}
  (h-m) \phi=0\,, \quad \bsd \phi=0\,,
\end{equation}
and the boundary condition $\Pj(\phi|_{z^\alpha=0})=\phi_0$,
where $\Pj$ is the projector to the subspace of totally traceless
elements in the space of $z^\alpha$ independent elements.
\end{prop}
The proof is completely similar to that of
Proposition~\bref{prop:hbsd-lift} given in Appendix~\bref{sec:C2}. The
proposition determines a map $K_m$ that sends a traceless
$z^\alpha$-independent element $\phi_0$ to a traceless element $\phi$
satisfying~\eqref{eq:K-def} and $\Pj(\phi|_{z^\alpha=0})=\phi_0$.

Furthermore, if $\phi=K_2\phi_0$ one finds that $\Pj((\sd\phi)|_{z^\alpha=0})=
\sd_0\phi_0$ where
\begin{equation}
  \sd_0=a^{\dagger b}\dl{Y^b}\,,\qquad h_0=a^{\dagger b}\dl{a^{\dagger
      b}}-y^{b}\dl{y^b}\,,\qquad
  \bsd_0=y^b\dl{a^{\dagger b}}\,,
\end{equation}
form the standard representation of $sl(2)$ in the space of
$z,w$-independent elements. It then follows that for any traceless
$z,w$-independent $\phi_0$ there exists a unique element
$\phi^\prime_0$ such that $\bsd_0 {\phi_0^\prime}=0$ and
$\phi^\prime_0=\phi_0+\sd_0 A_0$ for some traceless $z,w$-independent
element $A_0$. Indeed, in each irreducible component any element can
be uniquely represented as a linear combination of the element
annihilated by $\bsd_0$ (i.e., proportional to the lowest weight
vector) and the element in the image of $\sd_0$. Using
Proposition~\bref{prop:hbsd-lift-2} one then finds a unique $A$
satisfying $\bsd A=h A=0$ and $\Pj(A|_{z^\alpha=0})=A_0$ and finds
that $\phi^\prime=\phi+\sd A$ satisfies
$\Pj(\phi^\prime|_{z^\alpha=0})=\phi^\prime_0$.  Finally, one observes
that $\phi^\prime=\lift_m\phi^\prime_0$ does not depend on $w$ for any
$m$ provided $\bsd\phi_0=0$. Using
$\dl{w}=\frac{1}{l^2}V^A\dl{a^{\dagger A}}$ one then concludes that
the unique representative of a cohomology class in degree $0$ can be
assumed to satisfy
\begin{equation}
\label{eq:ce1-def}
  (h-2)\phi=0\,, \qquad \bsd \phi=0\,,\qquad V^A\dl{a^{\dagger A}}\phi=0\,.
\end{equation}
The decomposition of $\hat\cE$ then reads:
\begin{equation}
  \hat\cE=\cE\oplus\cF\oplus\cG\,,\qquad \cE=\cE_0\oplus\cE_1\,,
\end{equation}
where $\cE_1$ is the subspace of elements of the form $b^\dagger\chi$
with $\chi$ satisfying $h\phi=\sd\phi=\bsd\phi=0$, $\cE_0$ is
determined by \eqref{eq:ce1-def}, $\cG=\im \sd\dl{b^\dagger}$ in
$\hat\cE$, and $\cF$ is a complementary subspace.

We are now in the position to compute the reduced differential
\begin{equation}\label{eq:brst-unf}
  \brst^{\rm unf}=\st{\bundle\cE\bundle\cE}{D}
-\st{\bundle\cE\bundle\cF}{D}\rho\st{\bundle\cG\bundle\cE}{D}\,,\quad\qquad
D=\nabla+\sigma\,, 
\end{equation}
where $\rho\map \Gamma(\bundle\cG) \to \Gamma(\bundle\cF)$ is the
inverse to $\hat\brst_{-1}$. Note 
that because the cohomology is concentrated in degree $0$ and $1$,
the higher order terms in
  $(\st{\bundle\cG\bundle\cF}{\brst})^{-1}=\rho+\ldots$ cannot
  contribute and therefore there is only one additional term besides
$\st{\bundle\cE\bundle\cE}{\brst}$ in~\eqref{eq:brst-unf}.
\begin{prop}\label{prop:45}
  The system $(\hat\brst^\T,\Gamma(\bundle{\hat\cE}))$ can be
  consistently reduced to the unfolded system $(\brst^{\rm
    unf},\Gamma(\bundle{\cE}))$ with
\begin{equation}
\brst^{\rm unf}=\nabla+\sigma-b^\dagger
\sigma\bar\sigma\Pj_{R}\,,\qquad 
\bar\sigma=\theta^\mu e_\mu^A\dl{a^{\dagger A}}\,.
\end{equation}
Here $\Pj_{R}$ denotes the projector from $\cE_0$ to the space of
traceless elements described by rectangular Young tableaux defined as
follows: if $\phi=\phi^0+\phi^1+\ldots$ where $(a^{\dagger
  A}\dl{a^{\dagger A}}-Y^A\dl{Y^A})\phi^k=-k\phi^k$ then
$\Pj_{R}\phi=\phi^0$.
\end{prop}
Note that $\Pj_R$ is not a projector onto a subspace of $\cE_0$. The
proof of the proposition is again relegated to
Appendix~\bref{sec:C2}. Note that the last term in $\brst^{\rm unf}$
automatically belongs to $\cE_1$ because it follows from
$V^A\dl{a^{\dagger A}}\phi=0$ that $V^A\dl{a^{\dagger A}}\Pj_R \phi=0$
and therefore $(\bsd-V^A\dl{a^{\dagger A}})\Pj_R \phi=0$. At the same
time $(a^{\dagger A}\dl{a^{\dagger A}}-Y^A\dl{Y^A})\Pj_R \phi=0$.
Because $\bsd-V^A\dl{a^{\dagger A}}$, $h+V^A\dl{Y^A}$, and $\sd$ form
a standard presentation of $sl(2)$ on the space of $a^{\dagger A},
Y^A$-dependent elements, $\Pj_R\phi$ is $sl(2)$ invariant and
therefore $\sd \Pj_R\phi=0$.  Using $\bsd \Pj_R\phi=0$ one concludes
that $h\Pj_R\phi=0$ as well so that $b^\dagger
\sigma\bar\sigma\Pj_R\phi\in\cE_1$.

The equations of motion determined by $\brst^{\rm unf}$ take the form
$\brst^{\unf}\Psi^{\unf(0)}=0$ and explicitly read as
\begin{equation}
\label{eq:unf}
  (\nabla+\sigma) F=0\,,\qquad
(\nabla+\sigma)A+ \sigma\bar\sigma\Pj_R F=0\,,
\end{equation}
where the physical fields $A,F$ enter the zero-ghost-number component
of the string field:
\begin{equation}
  \Psi^{\unf(0)}=F(x;Y,a^\dagger)+b^\dagger\theta^\mu A_\mu(x;Y,a^\dagger)\,.
\end{equation}
Under a gauge transformation, $F$ is invariant while
$\delta A=(\nabla+\sigma)\lambda^{\unf}$ where
$\lambda^{\unf}(x;Y,a)$ is a gauge parameter with values in
the subspace of elements annihilated by $\sd,\bsd,h$.

\bigskip

Finally, we will rewrite the unfolded system in terms of
$z^\alpha$-independent fields in order to arrive at the formulation in
terms of intrinsic coordinates. To this end we assume that the local
frame is chosen such that $V^A=l\delta^A_{(d)}$ and as usually
$lw=a^{\dagger (d)}$, $lz=Y^{(d)}$, and $y^a=Y^a$.  It follows from
Proposition~\bref{prop:hbsd-lift-2} that any element of $\hat\cE$ is
uniquely determined by the traceless part of its
$z^\alpha$-independent part.  It is useful to describe subspaces
$\cE_0$ and $\cE_1$ in this way.  For $\cE_1$, equations
$h\chi=\bsd\chi=\sd\chi=0$ imply that $\chi_0=\Pj\chi|_{z^\alpha=0}$
satisfy $\sd_0\chi_0=0$ from which it follows that $\cE_1$ is
isomorphic to the space $\bar\cE_1$ of $z^\alpha$-independent and
linear in $b^\dagger$ traceless elements described by two row Young
tableaux for which the number of indexes contracted with $a^{\dagger
  b}$ is bigger than that contracted with $y^b$. The isomorphism is
just $\lift_0$ considered as a map from $\bar\cE_1$ to $\cE_1$.

For $\cE_0$, equations~\eqref{eq:ce1-def} imply that
$\phi_0=\phi|_{z^\alpha=0}$ satisfy $\bsd_0\phi_0=0$ and therefore the
space $\cE_0$ is isomorphic to the space $\bar\cE_0$ of
$b^\dagger,z^\alpha$-independent traceless elements described by two
row Young tableaux for which the number of indexes contracted with
$y^b$ is bigger than that contracted with $a^{\dagger b}$.  The
isomorphism is just $\lift_2$ considered as a map from $\bar\cE_0$ to
$\cE_0$.

This shows that the field content matches that of the unfolded form of
higher spin gauge fields in terms of intrinsic
coordinates~\cite{Vasiliev:2001wa,Lopatin:1988hz}.  It is instructive
to write down the structure of a representative $\phi$ of $\cE_0$
satisfying~\eqref{eq:ce1-def} in terms of its $z^\alpha$-independent
traceless part $\phi_0$. One finds
\begin{equation}
\label{eq:lift_2}
  \phi=\lift_2\phi_0=\frac{1}{(1+z)^{-h_0+2}}\left(
\phi_0+\frac{(n-s)(n-s+1)}{2l^2(d+2n-4)}(y^ay_a)\phi_0+\ldots
\right)
\end{equation}
where the ratio is understood as a formal power series and $\ldots$
denote terms proportional to $(y^ay_a)^k \phi_0$ with $k\geq 2$.

As for elements from $\cE_1$, let $\chi_0$ be a traceless
$z^\alpha$-independent element satisfying $\sd_0\chi_0=0$. Then its
representative in $\cE_1$ has the form
\begin{multline}
\chi=\lift_0 \chi_0=(z+1)^{h_0+N_w}\big(
\chi_0-w\bsd\chi_0+\ldots
+
\\
+(y^ay_a)(\ldots)
+(y^a a^\dagger_a)(\ldots)
+(a^{\dagger a}a^\dagger_a)(\ldots)\big)\,,
\end{multline}
where $N_w=w\dl{w}$, $\ldots$ denote terms proportional to $(w\bsd_0)^k\chi_0$
with $k\geq 2$, and terms in parenthesis denote some polynomials.

In addition we need the explicit expression for
$D=\nabla+\sigma$ in the frame where $V^A=l\delta^A_{(d)}$:
\begin{equation}
\label{eq:D}
  D=\nabla_0+\sigma-\frac{1}{l^2} e^a y_a\dl{z}
-\frac{1}{l^2} e^a a^\dagger_a \dl{\ww}
- e^a z\dl{y^a}
- e^a \ww \dl{a^{\dagger a}}\,,
\end{equation}
where $\nabla_0$ denotes the $d$-dimensional covariant derivative, i.e.,
\begin{equation}
  \nabla_0=\derham-\omega_{\,\, a}^by^a\dl{y^b}
-\omega_{\,\, a}^b a^{\dagger a}\dl{a^{\dagger b}}\,.
\end{equation}
\begin{prop}\label{prop:final}
  In terms of $z,w$ independent elements, the unfolded system takes
  the form
\begin{equation}
\label{eq:brst-unf-0}
\brst^\unf({\bundle\phi}_0+b^\dagger{\bundle\chi}_0)=D_{\bar\cE_0}{\bundle\phi}_0-b^\dagger D_{\bar\cE_1}{\bundle\chi}_0
-b^\dagger \sigma\bar\sigma\Pj^0_R {\bundle\phi}_0
\end{equation}
where ${\bundle\phi}_0\in \Gamma(\bundle{\bar\cE_0})$, $b^\dagger{\bundle\chi}_0\in \Gamma(\bundle{\bar\cE_1})$, and
$\Pj^0_R$ denotes the projector to the subspace of elements in
$\bar\cE_0$ described by rectangular Young tableaux. Furthermore, if
$n=y^a\dl{y^a}$ and $s=a^{\dagger a}\dl{a^{\dagger a}}$
\begin{multline}
\label{eq:DE0}
  D_{\bar\cE_0}{\bundle\phi}_0=
\nabla_0{\bundle\phi}_0+\sigma {\bundle\phi}_0+\frac{1}{n-s+2}\sd_0\bar\sigma{\bundle\phi}_0~+\\
+~\frac{(n-s+1)(d+n+s-4)}{l^2(d+2n-2)}\Pj\left[e_ay^a{\bundle\phi}_0 \right],
\end{multline}
and
\begin{multline}
\label{eq:DE1}
  D_{\bar\cE_1}{\bundle\chi}_0=
\nabla_0 {\bundle\chi}_0+\sigma{\bundle\chi}_0
-\frac{(d+s+n-4)}{l^2(d+2n-4)}\Pj\left[(s-n+1)e^a y_a
{\bundle\chi}_0-e^aa^\dagger_a\bsd_0{\bundle\chi}_0\right]\,.
\end{multline}
\end{prop}
Again, the proof is given in
Appendix~\bref{sec:C2}. Using~\eqref{eq:brst-unf-0},
equations~\eqref{eq:unf} take the form
\begin{equation}
  D_{\bar\cE_0}\bar F=0\,,\qquad D_{\bar\cE_1}\bar A+\sigma\bar\sigma
  \Pj^0_R \bar F=0\,.
\end{equation}
where $\bar F=\Pj(F|_{z^\alpha=0})$, $\bar A=\Pj(A|_{z^\alpha=0})$,
and $\Pj^0_R$ is the projector to the subspace of elements described
by rectangular Young tableaux. The fields entering $\bar F$ are gauge
invariant while for those in $\bar A$ one gets $\delta_{\bar\lambda}
\bar A=D_{\bar\cE_1}\bar\lambda$ where the gauge parameter takes
values in the subspace of $z^\alpha$-independent traceless elements
annihilated by $\sd_0$. Up to conventions and normalization factors,
these equations indeed coincide with the unfolded form of higher spin
equations~\cite{Lopatin:1988hz,Vasiliev:2001wa} in AdS space.

\section*{Acknowledgments}

\addcontentsline{toc}{section}{Acknowledgments}

The authors acknowledge useful discussions with G.~Bonelli.
M.G.~wants to thank the International Solvay Institutes for
hospitality. He is also grateful to K.~Alkalaev, I.~Tipunin,
A.~Semikhatov, I.~Tyutin, and M.~Vasiliev for useful discussions. The
work of G.B. is supported in part by a ``P\^{o}le d'Attraction
Interuniversitaire'' (Belgium), by IISN-Belgium, convention 4.4505.86,
by Proyectos FONDECYT 1970151 and 7960001 (Chile) and by the European
Commission program MRTN-CT-2004-005104, in which this author is
associated to V.U.~Brussels. The work of M.G.~was supported by the
RFBR Grant 04-01-00303 and by the Grant LSS-1578.2003.2.

\appendix

\section{Reduction in homological terms and
  $\cD$-modules}\label{sec:A1}

\begin{prop}\label{prop:red}
Suppose $\bundle\cH$ to be equipped with an additional grading besides the
ghost number,
\begin{equation}
  \bundle\cH=\bigoplus_{i\geq 0} \bundle\cH_{i},\qquad \deg(\bundle\cH_{i})=i,
\end{equation}
and let the BRST operator $\brst$ have the form
\begin{equation} \label{eq:2diffeq}
  \brst= \brst_{-1}
  +\brst_0
  +\sum_{i\geq1}\brst_i,\qquad \deg(\brst_i)=i,
\end{equation}
with $\brst_i:\Gamma(\bundle{\cH})_{j}\to\Gamma(\bundle{\cH})_{i+j}$.
If $\brst_{-1}$ is a linear map of vector bundles (i.e. does not
contain $x$-derivatives) then $H(\brst_{-1},\Gamma(\cH))\simeq
\Gamma(\bundle\cE)$ for some vector bundle $\bundle\cE$ and the system
$(\brst,\Gamma(\bundle{\cH}))$ can be consistently reduced to
$(\tilde\brst,\Gamma(\bundle{\cE}))$ where the operator $\tilde\brst$
is the differential induced by $\brst$ in the cohomology of
$\brst_{-1}$.
\end{prop}
Note that without loss of generality one can assume that $\bundle\cE$
is a subbundle in $\bundle\cH$. Moreover one can always find a
decomposition $\bundle\cH=\bundle\cE\oplus\bundle\cG\oplus\bundle\cF$
where $\Ker \brst_{-1}=\bundle\cE\oplus \bundle\cG$, $\bundle\cE\simeq
H(\brst_{-1},\bundle\cH)$, $\bundle\cG= {\rm Im}\,\brst_{-1}$, and
$\bundle\cF$ is a complementary subbundle. Then
$\st{\bundle\cG\bundle\cF}{\brst}$ is algebraically invertible and
$\tilde\brst$ is given by~\eqref{tilde-brst}.  Note also that if the
cohomology of $\brst_{-1}$ is concentrated in one degree then $\tilde
\brst=\brst_{0}$ considered as acting in $\Gamma(\bundle\cE)$.  Note
that Proposition~\bref{prop:red} is a slightly generalized version of
the one in~\cite{\BGST}.  After choosing an adapted local frame, its
proof reduces to that in~\cite{\BGST}.  In that reference, one can
also find an explicit recursive construction for $\tilde\brst$.

Let us note that two systems $(\brst,\Gamma(\bundle\cH))$ and
$(\brst^\prime,\Gamma(\bundle\cH^\prime))$ are obviously equivalent
when vector bundles $\bundle\cH$ and $\bundle\cH^\prime$ are
isomorphic and the isomorphism maps $\brst$ into $\brst^\prime$. At
the level of associated field theories the respective theories are
related by a field redefinition $\psi^A \to O^A_B(x)\psi^B $, where
$O^A_B$ are the components of the isomorphism map with respect to the
local frames. This is a very restricted class of field redefinitions
because it does not involve $x$-derivatives of fields.  A more general
class of equivalence relations is provided by allowing $O^A_B$ to be
algebraically invertible. Note that this notion of equivalence is
completely natural from the quantum mechanical point of view because
general similarity transformations in $x$-representation are allowed
to be invertible operators containing $x$-derivatives.

In fact there exists an adequate language which allows for a more
invariant formulation of first-quantized systems.  This amounts to
replacing the space of sections, which is a module over functions on
$\manX$, with the $\cD$-module over the algebra of differential
operators on $\manX$. We now shortly describe how to formulate the
basic notions in terms of $\cD$-modules.

Let $\cD_\manX$ be the algebra of differential operators on $\manX$ of
(graded) finite order.  Let also $\cD(\bundle\cH)$ be a right module
over $\cD_\manX$ generated by the vector bundle $\bundle\cH$.  Locally
on $\manX$, $\cD(\bundle\cH)$ is a free $\cD_\manX$-module generated
by a local frame $e_A$. A general element of $\cD(\bundle\cH)$ can
then be represented as $f=e_A f^A(x,\dl{x})$ where $e_A$ is a local
frame of $\bundle\cH$ and $f^A(x,\dl{x})$ are some differential
operators.  The action of linear differential operators on
$\bundle\cH$ can be extended to $\cD(\bundle\cH)$ as follows: if $G
(e_A{\bundle\phi}^A)=e_B G_A^B {\bundle\phi}^A$ for ${\bundle\phi} \in
\Gamma(\bundle\cH)$ then for $f\in\cD(\bundle\cH)$ one has $G f=G( e_A
f^A)=e_A G^A_B \circ f_B$ where $f\circ \phi$ denotes the composition
of differential operators, i.e., the associative product in the
algebra of operators on $\manX$.  Note that this left action is
compatible with the right module structure in the sense that it
commutes with the right multiplication by differential operators.

It is also natural to choose a local frame $e_A$ to be operator
valued.  For example, if $e_A$ is a local frame of $\bundle\cH$ and
$e^\prime_A=e_B O^B_A(x,\dl{x})$, with $O^A_B$ some algebraically
invertible differential operator, then in the new frame the BRST
operator $\brst$ takes the form $\brst e^\prime_A=e^\prime_B \circ
\brst^{\prime\,B}_A= e^\prime_B \circ (O^{-1})^B_C \circ \brst^C_D
\circ O^D_A$. If one associates fields $\psi^A$ and $\psi^{\prime A}$
to $e_A$ and $e^\prime_A$, the associated field theories determined by
$\brst$ and $\brst^\prime$ are related by a field redefinition
$\psi^A=O^A_B\psi^{\prime B}$.  In particular, two systems
$(\brst,\Gamma(\bundle\cH))$ and
$(\brst^\prime,\Gamma(\bundle\cH^\prime))$ are isomorphic if
$\cD(\bundle\cH^\prime)$ and $\cD(\bundle\cH))$ are isomorphic as
right $\cD_\manX$-modules and the isomorphism maps $\brst$ to
$\brst^\prime$.  We also note that, because the action of $\brst$
commutes with the right multiplication by a differential operators,
the kernel, image, cohomology etc. of $\brst$ are again
$\cD_\manX$-modules, though not necessarily generated by vector
bundles.

The approach just discussed is a BRST extension of the
standard $\cD$-module approach to partial differential equations (for
a review see e.g.~\cite{Schneiders}).
Note, however, that in the
standard approach, left $\cD$ modules are used which are in fact dual
to the ones described above.

\section{Particle on AdS: details of reductions}
\label{sec:B}
\subsection{Standard description}
\label{sec:B1}
Let us assume that we are in the frame where $V^A=l\delta^A_{(d)}$ and let
$z=lY^{(d)}$ and $y^a=Y^a$. Following subsection~4.2 of \cite{\BGST},
we choose as grading 
$Y^A\frac{\partial}{\partial
  Y^A}+2c_0\dl{c_0}$, so that the BRST operator \eqref{eq:2}
decomposes as $\brst
=\brst_{-1}+\brst_0$, where $\brst_{-1}=-\theta^\mu
e^a_{\mu}\frac{\partial}{\partial
  y^a}-\mu\frac{\partial}{\partial z}$, while
$$\brst_0=\theta^\mu(\dl{x^\mu}-\omega^A_{\mu\,B}Y^B\dl{Y^A})+
c_0\Box-\mu Y^A\dl{Y^A}+2c_0\eta \dl{c_0},$$
where $\Box{} =\dl{Y^A}\dl{Y_A}$.
It follows that
 \begin{equation}\label{H-EFG}
    \bundle{\cH}^{\T}=\bundle{\cE}\oplus\bundle{\cF}
\oplus\bundle{\cG}\nonumber
  \end{equation}
  where $\bundle{\cE}\cong H(\brst_{-1},\bundle{\cH}^{\T})$.
  Introducing $\rho=-y^a e_a^\mu\dl{\theta^\mu}-z\dl{\mu}$ and
  $N=Y^A\dl{Y^A}+\theta^\mu\dl{\theta^\mu}+\mu\dl{\mu}$, we choose
  $\bundle\cF=\rho\bundle{\cH}^{\T}$,
  $\bundle\cG=\brst_{-1}\bundle{\cH}^{\T}$. It then follows from
  Proposition~\bref{prop:red} that the system can be reduced to
  $(\tilde\brst,\Gamma(\bundle{\cE}))$. We still have to compute
\begin{eqnarray}
  \tilde\brst=\st{\bundle\cE\bundle\cE}{\brst}
  -\st{\bundle\cE\bundle\cF}{\brst}
  (\st{\bundle\cG\bundle\cF}{\brst})^{-1}\st{\bundle\cG\bundle\cE}{\brst}\,, 
\label{eq:18}
\end{eqnarray}
where in our case  $\st{\bundle\cE\bundle\cE}{\brst}=0$.
For this purpose, we introduce the
additional degree $\theta^\mu\dl{\theta^\mu}+\mu\dl{\mu}$, which we
denote by a superscript and according to which $\brst_0$ decomposes as
$\brst^0_0=c_0\Box$, while $
\brst^1_0=\theta^\mu(\dl{x^\mu}-\omega^A_{\mu\,B}Y^B\dl{Y^A})-\mu
Y^A\dl{Y^A}+2c_0\mu \dl{c_0}$. As in subsection~4.2 of \cite{\BGST},
we then get, for $\bundle{\phi}^{\bundle\cE}\in\Gamma(\bundle{\cE})$,
\begin{equation}
\begin{aligned}
  \label{eq:3b}
  \tilde\brst\bundle{\phi}^{\bundle\cE}&=\st{\bundle\cE\bundle\cF}{\brst^0_0}
\sum_{n=1}(-1)^n\big(N^{-1}\rho
\brst^1_0\big)^n\bundle{\phi}^{\bundle\cE}\\
&=\st{\bundle\cE\bundle\cF}{ \brst^0_0}\sum_{n=1}\frac{1}{n!}\big(
y^ae^\mu_a(\dl{x^\mu}-\omega^A_{\mu\, B}Y^B\dl{Y^A})-z( Y^A
\dl{Y^A}+2c_0\dl{c_0})\big)^n\bundle{\phi}^{\bundle\cE},
\\ &=
c_0\Box\half\big(y^ae^\mu_a(\dl{x^\mu}-\omega^A_{\mu\,
B}Y^B\dl{Y^A})-z Y^A
\dl{Y^A})(Y^be^\nu_b\ddl{\bundle{\phi}^{\bundle\cE}}{x^\nu})
\\ &= c_0e^{\mu a}(\delta^b_a \dl{x^\mu}-\omega^{b}_{\mu\,a})
(e^\nu_b\ddl{\bundle{\phi}^{\bundle\cE}}{x^\nu}) \\ &=
c_0\Box_{AdS}\bundle{\phi}^{\bundle\cE}.
\end{aligned}
\end{equation}

As an additional remark let us note that an alternative way to arrive
at the statement is to take as a degree minus the homogeneity in $\mu$
and $\theta^\nu$ so that
$\brst_{-1}=\nabla+\sigma+\mu(h-c_0\dl{c_0})$.  By expanding in $Y^A$
one finds that cohomology of $\brst_{-1}$ can be identified with
$\Gamma(\bundle\cE)$, i.e., $\theta^\mu,\mu,Y^A$-independent
sections. Using then a generalization of \bref{prop:red} discussed in
Appendix~\bref{sec:A1}, one can reduce the system to
$(\tilde\brst,\Gamma(\bundle\cE))$. Because the cohomology is
concentrated in zeroth degree, the reduced BRST operator $\tilde\brst$
is just $\brst_0=c_0\Box$ understood as acting in the cohomology and
coincides with~\eqref{eq:3b}.

\subsection{Unfolded form}
\label{sec:B2}
Because both $\brst_{-1}$ and the chosen degree (minus homogeneity in
$\mu$ and $c_0$) does not depend on the choice of frame, we are free
to use the frame where $V^A=l\delta^A_{(d)}$. The operator
$\brst_{-1}$ then has the form $\brst_{-1}= c_0\Box+\mu h-2\mu
c_0\dl{c_0}$ where
\begin{equation}
h=-y^a\dl{y^a}-(z+1)\dl{z}\,,\qquad \Box=\dl{y^a}\dl{y_a}-\frac{1}{l^2}
\dl{z}\dl{z}\,.
\end{equation}
One first evaluates the cohomology of $c_0\Box$. Because any element
is in the image of $\Box$, the cohomology is given by
$c_0$-independent elements annihilated by $\Box$. By expanding in
$c_0$, one concludes that the cohomology of $\brst_{-1}$ is given by
cohomology of $\mu h$ in the space of $c_0$-independent elements
annihilated by $\Box$.  Using the homogeneity degree in $z$ one then
concludes that the latter cohomology is determined by the cohomology
of $\mu\dl{z}$ in $\ker \Box$. We will now show that this cohomology
is given by $\mu,z$-independent elements in $\ker\Box$. Indeed, this
cohomology is isomorphic to the cohomology of $c_0\Box+\mu\dl{z}$ in
the space of all $\mu,z,c_0$-dependent elements: by expanding in $c_0$
one observes again that any cocycle can be assumed $c_0$-independent
and thus belonging to $\ker \Box$, while expanding in $\mu$, one
observes that any cocycle can be assumed $\mu,z$-independent. Finally,
the cohomology of $c_0\Box$ in the cohomology of $\mu\dl{z}$ is given
by $c_0,z,\mu$-independent elements from $\ker \Box$.

In degree $0$, the cohomology is $\cE={\rm Ker}\, h\cap {\rm Ker}\,
\Box$ and, by the above reasoning, this space is isomorphic to the
kernel of $\Box$ in the space of power series depending on $y^a=Y^a$
alone. We will now show how to uniquely lift such an element to a
$z$-dependent element that belongs to $\cE$. Any element in ${\rm
  Ker}\, h$ is uniquely determined by its $z$ independent
part: if this part vanishes, the element vanishes and, furthermore,
any element $\phi_0\in \ker \Box$ that does not depend on $z$ can be
completed to a unique solution of $h\phi=0$ and $\Box\phi=0$.  In fact
$\phi$ can be constructed order by order in $z$ using the fact that
any element in $\ker \Box$ is in the image of $\dl{z}$ in
$\ker\Box$. This is just a reformulation of the fact that $\mu\dl{z}$
does not have cohomology in non-vanishing degree in $\mu,z$.

The explicit expansion of $\phi$ in terms of $z$
and $y^ay_a$ has the following form:
\begin{equation}
  \phi=\frac{1}{(1+z)^n}(\phi_0+
  (y^ay_a)\frac{n(n+1)}{2l^2(d+2n)}\phi_0+\ldots)\,,
\end{equation}
where $n=y^a\dl{y^a}$, the ratio is understood as a formal power
series, and $\ldots$ denote terms of the form $(y^ay_a)^k \phi_0, \,\,
k\geq 2$. The coefficients in front of the terms $(y^ay_a)^k\phi_0$ are
uniquely determined by the requirement $\Box\phi=0$.

\section{Higher spin gauge fields on AdS: details of reductions}
\label{sec:C}
\subsection{Standard description}
\label{sec:C1}

To get the intermediate reduction to tensor fields taking values in
the embedding space, one chooses as a grading
$y^a\dl{y^a}+2c_0\dl{c_0}-b^\dagger\dl{b^\dagger}+c^\dagger\dl{c^\dagger}$.
The BRST operator \eqref{eq:intrinsic} decomposes as
\begin{align}
  \label{eq:7a}
&\brst_{-1}=-\theta^\mu(1+z)
  e^a_\mu\dl{y^a}\,,
\nonumber\\
&\brst_0=\theta^\mu\dl{x^\mu} -\theta^{\mu} (\omega^b_{\mu\,c}y^c
\dl{y^b}+\omega^B_{\mu\,C}
a^{\dagger C} \dl{a^{\dagger
B}})+c_0\dl{y_a}\dl{y^a} +a^{\dagger a}\dl{y^a}\dl{b^\dagger}+
\nonumber\\
&~+c^\dagger\dl{a^\dagger_a}\dl{y^a}
 +\xi T +\mu(-\dl{z}-Y^A\dl{Y^A}+a^{\dagger
A}\dl{a^{\dagger A}})+
\nu(\dl{w}+z\dl{w})-
\nonumber\\
&~-c^\dagger\dl{b^\dagger}\dl{c_0}-
2\xi\dl{b^\dagger}\dl{c^\dagger}  -2\mu c_0\dl{c_0}-2\mu
\dl{b^\dagger} b^\dagger +2\mu\xi\dl{\xi}+2\mu\nu\dl{\nu}
\,,
\\
&\brst_1=-\theta^\mu \frac{e^a_\mu}{l^2} y_a\dl{z}
+w\dl{z}\dl{b^\dagger}-\frac{c^\dagger}{l^2}\dl{w}\dl{z}
+\nu y^a\dl{a^{\dagger a}}
+2\nu c_0\dl{c^\dagger}
+\nu c^\dagger\dl{\xi}
+\nu\dl{b^\dagger}\dl{\mu}\,,
\nonumber\\
&\brst_2=-\frac{1}{l^2}c_0\dl{z}\dl{z}\,.
\nonumber
\end{align}
In this case, $\rho=-\frac{y^a}{1+z}e^\mu_a\dl{\theta^\mu}$ and
$N=y^a\dl{y^a}+\theta^\mu\dl{\theta^\mu}$, which determines the
decomposition
$\bundle{\cH^\T}=\bundle\cE\oplus\bundle\cG\oplus\bundle\cF$ with
$\Gamma(\bundle\cE)\cong H(\brst_{-1},\Gamma(\bundle{\cH^\T}))$ in the
same way as in~\bref{sec:B1}. The additional degree is
$\theta^\mu\dl{\theta^\mu}$, so that
  \begin{equation}
  \begin{aligned}
  \label{eq:11a}
&\brst^0_0=c_0\dl{y_a}\dl{y^a}
+a^{\dagger a}\dl{y^a}\dl{b^\dagger}+
\\
&~+c^\dagger\dl{a^\dagger_a}\dl{y^a} +\xi T
  +\mu(-\dl{z}-Y^A\dl{Y^A}+a^{\dagger A}\dl{a^{\dagger A}})+
  \nu(\dl{w}+z\dl{w})-
\\
&~-c^\dagger\dl{b^\dagger}\dl{c_0}-
  2\xi\dl{b^\dagger}\dl{c^\dagger} -2\mu c_0\dl{c_0}-2\mu
  \dl{b^\dagger} b^\dagger +2\mu\xi\dl{\xi}+2\mu\nu\dl{\nu} \,,
  \\
&\brst^0_1=
w\dl{z}\dl{b^\dagger}
-\frac{c^\dagger}{l^2}\dl{w}\dl{z}
+\nu y^a\dl{a^{\dagger a}}
+2\nu c_0\dl{c^\dagger}
+\nu c^\dagger\dl{\xi}
+\nu\dl{b^\dagger}\dl{\mu}\,,
  \\
&\brst^0_2=-\frac{1}{l^2}c_0\dl{z}\dl{z}\,,
  \\
&\brst^1_0=\theta^\mu\dl{x^\mu} -\theta^{\mu}
  (\omega^b_{\mu\,c}y^c \dl{y^b}+\omega^B_{\mu\,C}
  a^{\dagger C} \dl{a^{\dagger B}})\,,
\\
&\brst^1_1=-\theta^\mu \frac{e^a_\mu}{l^2} y_a\dl{z}\,.
\end{aligned}
\end{equation}
We now have
\begin{multline}
 \st{\bundle\cE\bundle\cE}{\brst}=
-\frac{1}{l^2}c_0\dl{z}\dl{z}
+w\dl{z}\dl{b^\dagger}
-\frac{c^\dagger}{l^2}\dl{w}\dl{z}
+\xi T
-
\\
-\mu(\dl{z}+
z\dl{z}-a^{\dagger A}\dl{a^{\dagger A}})+
\nu(\dl{w}+ z\dl{w})
-c^\dagger\dl{b^\dagger}\dl{c_0}+
\nu c^\dagger\dl{\xi}
+\nu\dl{b^\dagger}\dl{\mu}-
\\
-2\mu c_0\dl{c_0}-2\mu
\dl{b^\dagger} b^\dagger
+2\mu\xi\dl{\xi}+2\mu\nu\dl{\nu}-2\xi\dl{b^\dagger}\dl{c^\dagger}
+2\nu c_0\dl{c^\dagger}\,,\label{eq:17a}
\end{multline}
while
\begin{multline}
  \label{eq:3a}
  -\st{\bundle\cE\bundle\cF}\brst (\st{\bundle\cG\bundle\cF}\brst)^{-1}\st{\bundle\cG\bundle\cE}\brst\bundle{\phi^\cE}
~=~\st{\bundle\cE\bundle\cF}{\brst^0}
\sum_{n=1}(-1)^{n}\big(N^{-1}\rho(
\brst^1_0+\brst^1_1)\big)^n\bundle{\phi^\cE}=~\\
=~\st{\bundle\cE\bundle\cF}{ \brst^0}\sum_{n=1}
\frac{1}{n!}\Big(\frac{y^a}{1+z}\big(\partial_a-\omega^b_{a\,
c}y^c\dl{y^b}-\omega^B_{a\,C} a^{\dagger
C}\dl{a^{\dagger B}}-\frac{y_a}{2l^2}\dl{z}\big)\Big)^n\bundle{\phi^\cE}~=
\\ =~\Big[\frac{1}{1+z}(a^{\dagger a}
\cD_a\dl{b^\dagger}+c^\dagger
\dl{a^{\dagger}_a}\cD_a)
+
\\
+c_0[\big(\frac{1}{1+z}\big)^2
 \eta^{ac}(\delta^b_c\cD_a-\omega^b_{a\, c})\cD_b-\frac{1}{1+z}
\frac{\dmn}{l^2}\dl{z}]
\Big]{\bundle{\phi^\cE}}
\,,
\end{multline}
where $\partial_a=e^a_\mu\frac{\partial}{\partial x^\mu}$,
$\omega^B_{a\,C}=e^\mu_a\omega^B_{\mu\,C}$ and
$\cD_a=\partial_a-\omega^B_{a\,C}a^{\dagger
  C}\dl{a^{\dagger B}}$.
The reduced BRST operator is then given by \eqref{eq:12}.

To further reduce to tensor fields on AdS, we now choose as grading
$z\dl{z}+2c_0\dl{c_0}-b^\dagger\dl{b^\dagger}+c^\dagger\dl{c^\dagger}
-\nu\dl{\nu}$. The BRST operator \eqref{eq:12} then decomposes as
\begin{align}
  \label{eq:7}
  \brst_{-1}&=-\mu\dl{z}+\nu\dl{w}\,,
\nonumber\\
\brst_{\geq 0}&=-\frac{1}{l^2}c_0\dl{z}\dl{z}
+ w\dl{z}\dl{b^\dagger}
-\frac{c^\dagger}{l^2}\dl{w}\dl{z}
+\xi T
-\mu(z\dl{z}-a^{\dagger A}\dl{a^{\dagger A}})
+\nu z\dl{w}+
\nonumber\\
 &+\frac{1}{1+z}(a^{\dagger a}
\cD_a\dl{b^\dagger}
+c^\dagger\dl{a^{\dagger}_a}\cD_a)
+c_0[\big(\frac{1}{1+z}\big)^2
\eta^{ac}(\delta^b_c\cD_a
-\omega^b_{a\, c})\cD_b-\frac{1}{1+z}
\frac{\dmn}{l^2}\dl{z}]-
\nonumber\\
&-c^\dagger\dl{b^\dagger}\dl{c_0}
+\nu c^\dagger\dl{\xi}
+\nu\dl{b^\dagger}\dl{\mu}
-2\mu c_0\dl{c_0}
-2\mu\dl{b^\dagger} b^\dagger+
\nonumber\\ &
+2\mu\xi\dl{\xi}+2\mu\nu\dl{\nu}
-2\xi\dl{b^\dagger}\dl{c^\dagger}
+2\nu c_0\dl{c^\dagger}\,.
\end{align}
In this case, $\rho=-z\dl{\mu}+w\dl{\nu}$,
$N=z\dl{z}+w\dl{w} +\mu\dl{\mu}+\nu\dl{\nu}$. The
additional degree is $\mu\dl{\mu}+\nu\dl{\nu}$ so that
  \begin{align}
 \label{eq:11}
  \brst^0_{\geq 0}&= -\frac{1}{l^2}c_0\dl{z}\dl{z}
 +w\dl{z}\dl{b^\dagger} -\frac{c^\dagger}{l^2}\dl{w}\dl{z} +\xi T+
 \frac{1}{1+z}(a^{\dagger a} \cD_a\dl{b^\dagger}
 +c^\dagger\dl{a^{\dagger}_a}\cD_a)
\nonumber  \\
 &\qquad +c_0[\big(\frac{1}{1+z}\big)^2 \eta^{ac}(\delta^b_c\cD_a
 -\omega^b_{a\, c})\cD_b-\frac{1}{1+z} \frac{\dmn}{l^2}\dl{z}]-
\nonumber \\
 &\qquad -c^\dagger\dl{b^\dagger}\dl{c_0} +\nu\dl{b^\dagger}\dl{\mu}
 -2\xi\dl{b^\dagger}\dl{c^\dagger} \,,
\nonumber \\
  \brst^{1}_{\geq 0} &= \brst^{1}_{0}= -\mu(z\dl{z} -a^{\dagger
   A}\dl{a^{\dagger A}}) +\nu z\dl{w}+ +\nu c^\dagger\dl{\xi} -2\mu
 c_0\dl{c_0}
\nonumber \\
 &\qquad -2\mu \dl{b^\dagger} b^\dagger +2\mu\xi\dl{\xi}
 +2\mu\nu\dl{\nu} +2\nu c_0\dl{c^\dagger}\,.
\end{align}
We now have
\begin{multline}
\st{\bundle\cE\bundle\cE}{\brst}
=\xi \dl{a^{\dagger a}}\dl{a_a^\dagger}
+(c^\dagger\dl{a^{\dagger}_a}D_a
+a^{\dagger a} D_a\dl{b^\dagger})+
\\
+c_0[\eta^{ac}
(\delta^b_cD_a-\omega^b_{a\, c})D_b
+\frac{1}{l^2}a^{\dagger a}\dl{a^\dagger_a}]
-c^\dagger\dl{b^\dagger}\dl{c_0}
-2\xi\dl{b^\dagger}\dl{c^\dagger}
\,,\label{eq:17}
\end{multline}
where
\begin{equation}
D_a=\partial_a
-\omega^b_{a\,c}a^{\dagger c}\dl{a^{\dagger  b}}\,,
\quad
[D_a,D_b]=(\omega^c_{ab}-\omega^c_{ba})D_c
-\frac{1}{l^2}
(a^{\dagger}_a\dl{a^{\dagger b}}- a^{\dagger}_b\dl{a^{\dagger a}})\,,
\end{equation}
while
\begin{equation}
\begin{aligned}
  \label{eq:3}
  -\st{\bundle\cE\bundle\cF}\brst &(\st{\bundle\cG\bundle\cF}\brst)^{-1}\st{\bundle\cG\bundle\cE}\brst\bundle{\phi^\cE}
=\st{\bundle\cE\bundle\cF}{\brst^0}
\sum_{n=1}(-1)^{n}\big(N^{-1}\rho
\brst^1_{\geq 0}\big)^n\bundle{\phi^\cE}=
\\
&=\st{\bundle\cE\bundle\cF}{ \brst^0}\Big(z L
+w K+\half(zL+w K)^2-\half z^2 L+\dots\Big)\bundle{\phi^\cE}=
\\
&=\Big(\frac{c_0}{l^2}\big(L(1-\dmn-L)
-2Ka^{\dagger a}D_a +\frac{
K^2 a^{\dagger a}a^\dagger_a }{l^2}\big)-
\\
&\quad-\frac{c^\dagger}{l^2}\big(\half{\{L,K\}}
+K(\dmn +a^{\dagger a}\dl{a^{\dagger a}})\big)
-\frac{Ka^{\dagger a}a^{\dagger}_a}{l^2}\dl{b^{\dagger}}
-\frac{\xi}{l^2}K^2 \Big)\bundle{\phi^\cE}\,,
\end{aligned}
\end{equation}
where $\dots$ mean irrelevant terms of order at least $3$ in $w,z$,
while $L=a^{\dagger a}\dl{a^{\dagger
    a}}-2c_0\dl{c_0}-2\dl{b^\dagger}b^\dagger +2\xi\dl{\xi}$ and
$K=-c^\dagger\dl{\xi}-2c_0\dl{c^\dagger}$. By noting that $\half
\{K,L\}=(L+1)K$, $K^2=2c_0(1-2N_{c^\dagger})\dl{\xi}$ and using
\eqref{eq:17} and \eqref{eq:3} in the definition \eqref{tilde-brst},
we get the result \eqref{eq:16} of section~\bref{sec:4.3}.

\subsection{Unfolded form}
\label{sec:C2}
\paragraph{Proof of Proposition~\bref{prop:4inter}}

To prove that $H^n(\tilde\brst_{-1},\tilde\cE)=0$ for $n\neq 0$, let
us suppose that $V^A=l\delta^A_{(d)}$ and introduce the following
notations: $z^1=z=Y^{(d)}/l,z^2=w=a^{\dagger (d)}/l$ and
$\mu^1=\mu,\mu^2=\nu$.

By expanding in the homogeneity in $Y$ and $a^{\dagger}$, one finds
that the cohomology of $\tilde\brst_{-1}$ is controlled by the
cohomology of $\delta=\mu^\alpha\dl{z^\alpha}$.  As a preliminary
result, we need to show that the cohomology of $\delta$, which is
trivial in the space of formal power series in $Y$ with coefficients
that are polynomials in $a^{\dagger A},\mu^\alpha$ and the other ghost
variables, is also trivial in the space of traceless elements:
\begin{lemma}
  The cohomology of $\delta$ in $\tilde\cE$, the space of completely
  traceless elements, is given by $\mu^\alpha,z^\alpha$-independent
  traceless elements.
\end{lemma}
\begin{proof}
  The cohomology of $\delta$ in $\tilde\cE$ can be represented as that of
  $\brst_{\rm trace}+\delta$ in the space $\cH^\T$  where
  $\brst_{\rm trace}$ is given by~\eqref{eq:brst-trace}. Indeed,
  taking as a degree minus the homogeneity in ghosts
  $c_0,c^\dagger,\xi$ and using the fact that the cohomology of
  $\brst_{\rm trace}$ is concentrated in zeroth degree one concludes
  that the cohomology of $\brst_{\rm trace}+\delta$ is given by
  cohomology of $\delta$ in $\tilde \cE$.
  
  On the other hand, taking as a degree minus the homogeneity in
  $\mu^\alpha$ one finds that $\deg\delta=-1$, $\deg\brst_{\rm
    trace}=0$. The cohomology of $\delta$ is concentrated in zeroth
  degree and is given by $z^\alpha,\mu^\alpha$-independent elements. The
  cohomology of the entire operator is then given by that of
  $\brst_{\rm trace}$ reduced to the subspace of
  $z^\alpha,\mu^\alpha$-independent elements. The reduced differential
  is given by
\begin{equation}
  \begin{gathered}
    \brst_{\rm trace}^0=c_0\Box_0+c^\dagger T_0+\xi S_0\,, \\
    \Box_0=\frac{\d^2}{\d y^a\d y_a}\,,\qquad T_0=\frac{\d^2}{\d
      a^{\dagger a}\d a^{\dagger}_a}\,,\qquad S_0=\frac{\d^2}{\d y^a\d
      a^{\dagger}_a}\,.
\end{gathered}
\end{equation}
According to the results of~\cite{\BGST}, the cohomology of $\brst^0_{\rm
trace}$ is given by $c_0,c^\dagger,\xi$-independent traceless elements provided
$d\geq 3$.
\end{proof}
Analogous arguments show that the cohomology of
$\delta^1=\mu^1\dl{z^1}$ (respectively $\delta^2=\mu^2\dl{z^2}$) in
the space of traceless elements is given by $\mu^1,z^1$- (respectively
$\mu^2,z^2$) independent traceless elements. Let $\tilde\cE_0\subset
\tilde \cE$ be the subspace of $\mu^\alpha,z^\alpha$-independent
elements. One then has the following:
\begin{prop}\label{prop:hbsd-lift}
  For any $\phi\in\tilde\cE_0$ satisfying $\bsd \phi=0$ and an
  arbitrary number $m$, there exists a unique $z^\alpha$-dependent and
  $\mu^\alpha$-independent $A\in \tilde\cE$ such that
\begin{equation}\label{eq:1-prop}
  \phi= (h-m) A\,, \quad \bsd A=0\,, \quad \Pj(A|_{z^\alpha=0})=0\,,
\end{equation}
where $\Pj$ is the projector to the subspace of totally traceless
elements in the space of $z^\alpha$ independent elements.
\end{prop}
\begin{proof}
  Multiplying equations~\eqref{eq:1-prop} respectively by $\mu^1$ and
  $\mu^2$, one gets
\begin{equation}
  \delta A=\Delta A-\mu^1 \phi\,,\qquad \Delta=\mu^1
  (h-m+\dl{z^1})-\mu^2(\bsd-\dl{z^2})\,,\quad
  \delta=\mu^\alpha\dl{z^\alpha}\,.
\end{equation}
Note that $\Delta$ is homogeneous of degree $0$ in the total degree
that counts $Y$'s and $a^\dagger$'s. By expanding according to this
total degree, one gets the equations
\begin{equation}
  \delta A_{n+1}=\Delta A_n-\mu^1\phi_n\,,
\end{equation}
which has a unique $\mu^\alpha$-independent solution satisfying the
condition $\Pj(A_{n}|_{z^\alpha=0})=0$ for all $n$. This follows from
the fact that the cohomology of $\delta$ is given by $z^\alpha$ and
$\mu^\alpha$ independent elements while the consistency holds due to
the following identities
\begin{equation}
  \commut{\delta}{\Delta}=-2\mu^1\delta\,,\qquad
  \half\commut{\Delta}{\Delta}=-2\mu^1\Delta\,,
\end{equation}
and $\bsd \phi=0$. In order to see that the solution
is unique one notes that the arbitrariness in $A_{n+1}$ is given by
$z^\alpha$-independent terms which are traceless. By requiring the
traceless part of $A|_{z^\alpha=0}$ to vanish one thus fixes the ambiguity.
Note that in general $A$ is a formal power series in $z^1$ even if
$\phi$ is polynomial in all the variables.
\end{proof}
Similar arguments show that any element from $\tilde\cE_0$ is in the
image of $\bsd$. One then has all the ingredients needed in the proof
of Proposition~\bref{prop:4inter}.

\paragraph{Proof of Proposition~\bref{prop:45}}
First, one observes that
the second term in \eqref{eq:brst-unf} vanishes when acting on a
$\cE_1$-valued section and therefore one gets
\begin{equation}
  \brst^{\rm unf}\bundle\chi=(\nabla+\sigma) \bundle\chi\,,\qquad \bundle\chi\in
  \Gamma(\bundle{\cE_1})\,.
\end{equation}
Note that the projection here is not needed because $D=\nabla+\sigma$
preserves $\Gamma(\bundle{\cE_1})$.

To compute $\brst^{\unf}\bundle\phi$ for
$\bundle\phi\in\Gamma(\bundle\cE_0)$ it is convenient to choose the
frame where $V^A=l\delta^A_{(d)}$. One is allowed to use a special
frame because the statement is frame-independent.  As usually we also
use $lz=lz^1=Y^{(d)}$, $lw=lz^2=a^{\dagger (d)}$. For a $\cE_0$-valued
section again $\st{\bundle\cE\bundle\cE}{D}\bundle\phi=D\bundle\phi$
but the second term in \eqref{eq:brst-unf} can be non-vanishing. To
compute it explicitly, one first observes that it can be non-vanishing
only on elements whose traceless part at $z^\alpha=0$ is annihilated
by $h_0$, i.e., is described by a rectangular Young tableaux. This
follows from counting homogeneity degree in $y^a,a^{\dagger a}$.  Let
$\bundle\phi\in\Gamma(\bundle\cE_0)$ be such that the traceless part
$\bundle\phi_0$ of $\bundle\phi|_{z^\alpha=0}$ satisfies
$h\bundle\phi_0=h_0\bundle\phi_0=0$. One then observes that
\begin{equation}
  \Pj_\cG D\bundle\phi=-\sd \lift_0 (\bar\sigma \bundle\phi_0)\,,\qquad
  \bar\sigma=-\theta^\mu e_\mu^A\dl{a^{\dagger A}}\,,
\end{equation}
where $\bundle X=\lift_0 ( \bar\sigma \bundle\phi_0)$ is a unique
solution $\bundle{X}\in\Gamma(\bundle{\hat\cE})$ 
to the equation $h \bundle{X}=\bsd \bundle{X}=0$ satisfying
$\Pj (\bundle{X}|_{z^\alpha=0})=\bar\sigma\bundle\phi_0$ (see
Proposition~\bref{prop:hbsd-lift-2}).  Indeed, to see this it is
enough to show that $D\bundle\phi+\sd \bundle{X}\in
\Gamma(\bundle{\cE_0})$ which is in turn equivalent to 
$\dl{\ww}(D\bundle\phi+\sd \bundle{X})=0$.
One then uses $\dl{w}\bundle{X}=\dl{w}\bundle\phi=0$, $\commut{\dl{\ww}}{D}=\bar\sigma$,
$\commut{\dl{\ww}}{\sd}=\dl{z}$ and finds that both $\dl{z}\bundle{X}$ and
$-\bar\sigma\bundle\phi$ are annihilated by $h-1$ and $\bsd$ and satisfy the
same boundary condition at $z=0$.  It then follows from the
Proposition~\bref{prop:hbsd-lift-2} that they do coincide.

By counting degree in $y^b$ and $a^{\dagger b}$ one finds that
$\Pj_{\cE_1} b^\dagger \lift_0(\bar\sigma \bundle\phi_0)=0$ because
$b^\dagger \bar\sigma\bundle\phi_0$ does not belong to $\bar\cE_1$. Thus
\begin{equation}
  \rho \st{\cG\cE_0}{D}\bundle\phi=
  -b^\dagger\lift_0(\bar\sigma \bundle\phi_0)\,.
\end{equation}
Again, by counting the degree one finds that the only contribution to
$\cE_1$ from $-Db^\dagger\lift_0(\bar\sigma \bundle\phi_0)$ comes from the
terms in $D$ that lower the degree in $y^a$. These terms give
$(z+1)\sigma$ and one finds that
\begin{equation}
\label{eq:contr}
  \Pj_{\cE_1}b^\dagger D \lift_0(\bar\sigma \bundle\phi_0)=
\lift_0(b^\dagger\sigma\bar\sigma\bundle\phi_0)=b^\dagger
\sigma\bar\sigma\bundle\phi_0 
\end{equation}
where we have used $\commut{h}{(z+1)\sigma}=0$,
$\commut{\bsd}{(z+1)\sigma}=-(z+1)\bar\sigma$, and
$\bar\sigma\lift(\bar\sigma\bundle\phi_0)=0$. The last equality
follows from the fact that
$h_0\sigma\bar\sigma\bundle\phi_0=\bsd_0\sigma\bar\sigma\bundle\phi_0
=\sd_0\sigma\bar\sigma\bundle\phi_0=0$ and therefore $\lift_0
\sigma\bar\sigma\bundle\phi_0=\sigma\bar\sigma\bundle\phi_0$.

There remains to show that for arbitrary
$\bundle\phi\in\Gamma(\bundle\cE_0)$ one has $\Pj_R
\bundle\phi=\Pj^0_R\bundle\phi_0$ where $\Pj^0_R$ denotes the standard
projector onto the subspace of elements in $\bar\cE_0$ described by
rectangular Young tableaux. Indeed, it follows from
$\bsd_0\bundle\phi_0=0$ and the explicit structure~\eqref{eq:lift_2}
of $\bundle\phi=\lift_2 \bundle\phi_0$ that all monomials in
$\bundle\phi-\bundle\phi_0$ contain more $Y^A$ variables than
$a^{\dagger A}$ ones. This allows to rewrite the
contribution~\eqref{eq:contr} in frame-independent terms
\begin{equation}
  b^\dagger \sigma\bar\sigma\Pj^0_R\bundle\phi_0=  b^\dagger
  \sigma\bar\sigma\Pj_R\bundle\phi\,, 
\end{equation}
where $\bundle\phi_0=\Pj(\bundle\phi|_{z^\alpha=0})$ and
$\bundle\phi\in\cE_0$.

\paragraph{Proof of Proposition~\bref{prop:final}}

First one observes that $D_{\bar\cE_0}{\bundle\phi}_0=\Pj_{\bar\cE_0}
\Pj[(D{\bundle\phi})|_{z^\alpha=0}]$ where
${\bundle\phi}\in\Gamma(\bundle{\cE_0})$ satisfies
$\Pj[{\bundle\phi}|_{z^\alpha=0}]={\bundle\phi}_0$.  To compute
$D_{\bar\cE_0}{\bundle\phi}_0$ explicitly one then observes that the
first two terms in~\eqref{eq:D} obviously commute with putting
$z^\alpha$ to zero as well as with the projection to the subspace of
elements annihilated by $\bsd_0$. From the rest of the terms the third
and the fourth ones give some additional contributions.

In order to find $\Pj_{\bar\cE_0}\Pj[(\sigma{\bundle\phi})|_{z^\alpha=0}]$ one
notices that it follows from $(h-2){\bundle\phi}=0$ that the expansion of
${\bundle\phi}$ in $z$ has the form
\begin{equation}
\label{eq:z-exp}
  {\bundle\phi}=\frac{1}{(z+1)^{n-s+2}}{\bundle\phi}|_{z=0}\,,
\end{equation}
where $n=y^a\dl{y^a}$ and $s=a^{\dagger a}\dl{a^{\dagger a}}$ are the
operators counting the degree of homogeneity in $y^a$ and $a^{\dagger
  a}$. On the other hand, taking a $\Box$-traceless part of
$\Box{\bundle\phi}=0$ at $z=0$ and using
$\Box_z=-\frac{1}{l^2}\dl{z}\dl{z}$, one finds
\begin{equation}
\label{eq:y2-exp}
  {\bundle\phi}|_{z=0}={\bundle\phi}_0
+\frac{(n-s)(n-s+1)}{2l^2(d+2n-4)}(y^ay_a){\bundle\phi}_0+\ldots
\end{equation}
where $\ldots$ denote terms proportional to higher powers in $y^ay_a$.
Computing the projector explicitly one then finds
\begin{multline}
  \Pj_{\bar\cE_0}(\sigma{\bundle\phi}|_{z^\alpha=0})=\sigma{\bundle\phi}_0
+\frac{1}{n-s+1}\sd_0\bar\sigma{\bundle\phi}_0
~-\\
-~\frac{(n-s+1)(n-s+2)}{l^2(d+2n-2)}\Pj\left[(e^ay_a){\bundle\phi}_0\right]\,.
\end{multline}

It follows from the expansions~\eqref{eq:z-exp} and \eqref{eq:y2-exp}
that
\begin{equation}
  \Pj[(\dl{z}{\bundle\phi})_{z^\alpha=0}]=-(n-s+2){\bundle\phi}_0\,,
\end{equation}
which in turn implies
\begin{equation}
  \Pj_{\bar\cE_0}\Pj\left[(-\frac{1}{l^2}
e^a y_a \dl{z}{\bundle\phi})|_{z^\alpha=0}\right]=
\frac{n-s+1}{l^2}\Pj\left[ e^a y_a {\bundle\phi}_0 \right]\,.
\end{equation}
Note that here the projection $\Pj_{\bar\cE_0}$ is omitted because
$\bsd_0$ commutes with $e^ay_a$ so that the result belongs to
$\bar\cE_0$ automatically. Summing up all contribution one arrives
at~\eqref{eq:DE0}.

For $D_{\bar\cE_1}$ one finds
$D_{\bar\cE_1}{\bundle\chi}_0=\Pj_{\bar\cE_1}\Pj[(D{\bundle\chi})|_{z^\alpha=0}]$ where
${\bundle\chi}$ is uniquely determined by $b^\dagger {\bundle\chi}\in\Gamma(\bundle{\cE_1})$ and
$\Pj[{\bundle\chi}|_{z^\alpha=0}]={\bundle\chi}_0$.  In this case one observes that
only the first five terms in~\eqref{eq:D} do contribute to
$\Pj_{\bar\cE_1}\Pj[(D{\bundle\chi})|_{z^\alpha=0}]$. Finding the explicit
expression is a bit more difficult in this case. The relevant terms in
the expansion of ${\bundle\chi}$ in terms of $z,\ww$ and traces read as:
\begin{multline}
{\bundle\chi}={\bundle\chi}_0+z(s-n){\bundle\chi}_0-\ww\bsd_0{\bundle\chi}_0
+
\\
+\half(s-n)(s-n+1)zz{\bundle\chi}_0-
(s-n+1)z\ww\bsd_0{\bundle\chi}_0+\half ww\bsd_0\bsd_0{\bundle\chi}_0+\ldots+
\\
+
\alpha (y^a y_a){\bundle\chi}_0 + \beta
(y^aa^\dagger_a)\bsd_0{\bundle\chi}_0+\gamma(a^{\dagger
  a}a^\dagger_a)\bsd_0\bsd_0{\bundle\chi}_0
+\ldots
\end{multline}
where $\alpha,\beta,\gamma$ are coefficients depending on $n,s$. It is
useful to assume for the moment that ${\bundle\chi}_0$ is homogeneous so that
$n{\bundle\chi}_0=\nn$ and $s{\bundle\chi}_0=\ss$. The tracelessness condition then
implies:
\begin{equation}
  \begin{split}
   - \frac{1}{l^2}(\ss-\nn)(\ss-\nn-1)+2\alpha(d+2\nn)+2\beta(\ss-\nn)&=0\,,\\
\frac{1}{l^2}(\ss-\nn-1)+2\alpha+\beta(d+\nn+\ss)+4\gamma (\ss-\nn-1)&=0\,,\\
-\frac{1}{l^2}+2\beta+2\gamma(d+2\ss-4)&=0\,.
  \end{split}
\end{equation}
which determines the coefficients to be
\begin{equation}
  \alpha=\frac{(\ss-\nn)(\ss-\nn-1)}{2l^2(d+2\nn-2)}\,,\quad
\beta=-\frac{\ss-\nn-1}{l^2(d+2\nn-2)}\,,\quad
\gamma=\frac{1}{2l^2(d+2\nn-2)}.
\end{equation}

Now one can explicitly compute all contributions:
\begin{multline}
  \Pj_{\bar\cE_1}\Pj\left[(\sigma{\bundle\chi})|_{z^\alpha=0}\right]=\sigma{\bundle\chi}_0
  -{2\alpha}
  \Pj_{\bar\cE_1}\Pj[e^ay_a{\bundle\chi}_0]-{\beta}\Pj_{\bar\cE_1}\Pj[e^a
  a^\dagger_a
  \bsd_0 {\bundle\chi}_0]~=
  \\
  =~(-2\alpha+\beta)\Pj_{\bar\cE_1}\Pj[e^ay_a{\bundle\chi}_0]~=
\\=~-\frac{(s-n)}{l^2(d+2n-4)}\Pj[(s-n+1)e^ay_a{\bundle\chi}_0-e^a
  a^\dagger_a \bsd_0 {\bundle\chi}_0]\,,
\end{multline}
where we have re-expressed the coefficients in terms of $n,s$. Note that the
projection to the subspace of elements annihilated by $\sd_0$ is
automatic in the last expression.  Furthermore,
\begin{equation}
  \Pj_{\bar\cE_1}\Pj\left[-\frac{1}{l^2} e^ay_a(\dl{z}{\bundle\chi})|_{z^\alpha=0}\right]=
-\frac{s-n+1}{l^2}\Pj_{\bar\cE_1}\Pj\left[e^ay_a{\bundle\chi}_0\right]
\end{equation}
and
\begin{equation}
  \Pj_{\bar\cE_1}\Pj\left[-\frac{1}{l^2}
    e^aa^\dagger_a(\dl{\ww}{\bundle\chi})|_{z^\alpha=0}\right]=
\frac{1}{l^2}\Pj_{\bar\cE_1}\Pj\left[e^aa^\dagger_a\bsd_0{\bundle\chi}_0\right]\,.
\end{equation}
Summing up all contribution one arrives at~\eqref{eq:DE1}.

Finally, the structure of the last term in \eqref{eq:brst-unf-0} is
obvious if one observes that, for ${\bundle\phi}_0\in\Gamma(\bundle{\bar\cE_0})$ and
${\bundle\phi}=\lift_2{\bundle\phi}_0$, the projectors coincide:
$\Pj^0_R{\bundle\phi}_0=\Pj_R{\bundle\phi}$ and
$\lift_0\sigma\bar\sigma\Pj^0_R{\bundle\phi}_0=\sigma\bar\sigma\Pj^0_R{\bundle\phi}_0$.

\addcontentsline{toc}{section}{References}

\providecommand{\href}[2]{#2}\begingroup\raggedright\endgroup

\begin{thebibliography}{10}
\addtolength{\parskip}{-3pt}
\addtolength{\baselineskip}{-2pt}


\bibitem{Fierz:1939ix}
M.~Fierz and W.~Pauli, ``On relativistic wave equations for particles of
  arbitrary spin in an electromagnetic field,'' {\em Proc. Roy. Soc. Lond.}
  {\bf A173} (1939)
211--232.
%%CITATION = PRSLA,A173,211;%%.

\bibitem{Singh:1974qz}
L.~P.~S. Singh and C.~R. Hagen, ``Lagrangian formulation for arbitrary spin. 1.
  {T}he boson case,'' {\em Phys. Rev.} {\bf D9} (1974)
898--909.
%%CITATION = PHRVA,D9,898;%%.

\bibitem{Fronsdal:1978rb}
C.~Fronsdal, ``Massless fields with integer spin,'' {\em Phys. Rev.} {\bf D18}
  (1978)
3624.
%%CITATION = PHRVA,D18,3624;%%.

\bibitem{Fronsdal:1979vb}
C.~Fronsdal, ``Singletons and massless, integral spin fields on de {S}itter
  space (elementary particles in a curved space {VII}),'' {\em Phys. Rev.} {\bf
  D20} (1979)
848--856.
%%CITATION = PHRVA,D20,848;%%.

\bibitem{Siegel:1984ap}
W.~Siegel, ``Covariantly second quantized string,'' {\em Phys. Lett.} {\bf
  B142} (1984)
276.
%%CITATION = PHLTA,B142,276;%%.

\bibitem{Siegel:1984wx}
W.~Siegel, ``Covariantly second quantized string. 2,'' {\em Phys. Lett.} {\bf
  B149, B151} (1984, 1985)
157,391.
%%CITATION = PHLTA,B149,157;%%.

\bibitem{Siegel:1984xd}
W.~Siegel, ``Covariantly second quantized string. 3,'' {\em Phys. Lett.} {\bf
  B149, B151} (1984, 1985)
162, 396.
%%CITATION = PHLTA,B149,162;%%.

\bibitem{Siegel:1985tw}
W.~Siegel and B.~Zwiebach, ``Gauge string fields,'' {\em Nucl. Phys.} {\bf
  B263} (1986)
105.
%%CITATION = NUPHA,B263,105;%%.

\bibitem{Neveu:1985ya}
A.~Neveu and P.~C. West, ``Gauge covariant local formulation of bosonic
  strings,'' {\em Nucl. Phys.} {\bf B268} (1986)
125.
%%CITATION = NUPHA,B268,125;%%.

\bibitem{Neveu:1985cx}
A.~Neveu, H.~Nicolai, and P.~C. West, ``Gauge covariant local formulation of
  free strings and superstrings,'' {\em Nucl. Phys.} {\bf B264} (1986)
573.
%%CITATION = NUPHA,B264,573;%%.

\bibitem{Witten:1985cc}
E.~Witten, ``Noncommutative geometry and string field theory,'' {\em Nucl.
  Phys.} {\bf B268} (1986)
253.
%%CITATION = NUPHA,B268,253;%%.

\bibitem{Ohta:1985zw}
N.~Ohta, ``Covariant second quantization of superstrings,'' {\em Phys. Rev.
  Lett.} {\bf 56} (1986)
440 [Erratum--ibid.\ {\bf 56}, 1316 (1986)].
%%CITATION = PRLTA,56,440;%%.

\bibitem{deAlwis:1986mq}
S.~P. de~Alwis and N.~Ohta, ``All free string theories are theories of {BRST}
  cohomology,'' {\em Phys. Lett.} {\bf B174} (1986)
388.
%%CITATION = PHLTA,B174,388;%%.

\bibitem{Thorn:1989hm}
C.~B. Thorn, ``String field theory,'' {\em Phys. Rept.} {\bf 175} (1989)
1--101.
%%CITATION = PRPLC,175,1;%%.

\bibitem{Ouvry:1986dv}
S.~Ouvry and J.~Stern, ``Gauge fields of any spin and symmetry,'' {\em Phys.
  Lett.} {\bf B177} (1986)
335.
%%CITATION = PHLTA,B177,335;%%.

\bibitem{Bengtsson:1986ys}
A.~K.~H. Bengtsson, ``A unified action for higher spin gauge bosons from
  covariant string theory,'' {\em Phys. Lett.} {\bf B182} (1986)
321.
%%CITATION = PHLTA,B182,321;%%.

\bibitem{Henneaux:1987cp}
M.~Henneaux and C.~Teitelboim, {\em First and second quantized point particles
  of any spin}, ch.~9, pp.~113--152.
\newblock Quantum mechanics of fundamental systems 2, Centro de Estudios
  Cient\'{\i}ficos de Santiago.
\newblock Plenum Press, 1987.

\bibitem{Bouatta:2004kk}
N.~Bouatta, G.~Compere, and A.~Sagnotti, ``An introduction to free higher-spin
  fields,''
\href{http://www.arXiv.org/abs/hep-th/0409068}{{\tt hep-th/0409068}}.
%%CITATION = HEP-TH 0409068;%%.

\bibitem{Vasiliev:2003ev}
M.~A. Vasiliev, ``{N}onlinear equations for symmetric massless higher spin
  fields in (a)d{S}(d),'' {\em Phys. Lett.} {\bf B567} (2003) 139--151,
\href{http://www.arXiv.org/abs/hep-th/0304049}{{\tt hep-th/0304049}}.
%%CITATION = HEP-TH 0304049;%%.

\bibitem{Vasiliev:1990en}
M.~A. Vasiliev, ``Consistent equation for interacting gauge fields of all spins
  in (3+1)-dimensions,'' {\em Phys. Lett.} {\bf B243} (1990)
378--382.
%%CITATION = PHLTA,B243,378;%%.

\bibitem{Vasiliev:1988xc}
M.~A. Vasiliev, ``Equations of motion of interacting massless fields of all
  spins as a free differential algebra,'' {\em Phys. Lett.} {\bf B209} (1988)
491--497.
%%CITATION = PHLTA,B209,491;%%.

\bibitem{Vasiliev:1988sa}
M.~A. Vasiliev, ``Consistent equations for interacting massless fields of all
  spins in the first order in curvatures,'' {\em Annals Phys.} {\bf 190} (1989)
59--106.
%%CITATION = APNYA,190,59;%%.

\bibitem{Vasiliev:2001zy}
M.~A. Vasiliev, ``Conformal higher spin symmetries of {4D} massless
  supermultiplets and {osp(L,2M)} invariant equations in generalized
  (super)space,'' {\em Phys. Rev.} {\bf D66} (2002) 066006,
\href{http://www.arXiv.org/abs/hep-th/0106149}{{\tt hep-th/0106149}}.
%%CITATION = HEP-TH 0106149;%%.

\bibitem{Alkalaev:2003qv}
K.~B. Alkalaev, O.~V. Shaynkman, and M.~A. Vasiliev, ``On the frame-like
  formulation of mixed-symmetry massless fields in (a)ds(d),'' {\em Nucl.
  Phys.} {\bf B692} (2004) 363--393,
\href{http://www.arXiv.org/abs/hep-th/0311164}{{\tt hep-th/0311164}}.
%%CITATION = HEP-TH 0311164;%%.

\bibitem{Sezgin:2001ij}
E.~Sezgin and P.~Sundell, ``{7D} bosonic higher spin theory: {S}ymmetry algebra
  and linearized constraints,'' {\em Nucl. Phys.} {\bf B634} (2002) 120--140,
\href{http://www.arXiv.org/abs/hep-th/0112100}{{\tt hep-th/0112100}}.
%%CITATION = HEP-TH 0112100;%%.

\bibitem{Barnich:2004cr}
G.~Barnich, M.~Grigoriev, A.~Semikhatov, and I.~Tipunin, ``Parent field theory
  and unfolding in {BRST} first-quantized terms,'' {\em Commun. Math. Phys.}
  {\bf 260} (2005) 147--181,
\href{http://www.arXiv.org/abs/hep-th/0406192}{{\tt hep-th/0406192}}.
%%CITATION = HEP-TH 0406192;%%.

\bibitem{Vasiliev:2001wa}
M.~A. Vasiliev, ``Cubic interactions of bosonic higher spin gauge fields in
  {AdS}(5),'' {\em Nucl. Phys.} {\bf B616} (2001) 106--162 [Erratum--ibid. B
  {\bf 652}, 407 (2003)],
\href{http://www.arXiv.org/abs/hep-th/0106200}{{\tt hep-th/0106200}}.
%%CITATION = HEP-TH 0106200;%%.

\bibitem{MacDowell:1977jt}
S.~W. MacDowell and F.~Mansouri, ``Unified geometric theory of gravity and
  supergravity,'' {\em Phys. Rev. Lett.} {\bf 38} (1977)
739.
%%CITATION = PRLTA,38,739;%%.

\bibitem{Stelle:1979aj}
K.~S. Stelle and P.~C. West, ``Spontaneously broken de {S}itter symmetry and
  the gravitational holonomy group,'' {\em Phys. Rev.} {\bf D21} (1980)
1466.
%%CITATION = PHRVA,D21,1466;%%.

\bibitem{Bekaert:2005vh}
X.~Bekaert, S.~Cnockaert, C.~Iazeolla, and M.~A. Vasiliev, ``Nonlinear higher
  spin theories in various dimensions,''
\href{http://www.arXiv.org/abs/hep-th/0503128}{{\tt hep-th/0503128}}.
%%CITATION = HEP-TH 0503128;%%.

\bibitem{Fedosov-book}
B.~Fedosov, ``Deformation quantization and index theory,''. Berlin, Germany:
  Akademie-Verl. (1996) 325 p. (Mathematical topics: 9).

\bibitem{Bordemann:1997er}
M.~Bordemann, N.~Neumaier, and S.~Waldmann, ``Homogeneous {F}edosov star
  products on cotangent bundles {II}: {GNS} representations, the {WKB}
  expansion, and applications,''
\href{http://www.arXiv.org/abs/q-alg/9711016}{{\tt q-alg/9711016}}.
%%CITATION = Q-ALG 9711016;%%.

\bibitem{Batalin:1989mb}
I.~A. Batalin, E.~S. Fradkin, and T.~E. Fradkina, ``Generalized canonical
  quantization of dynamical systems with constraints and curved phase space,''
  {\em Nucl. Phys.} {\bf B332} (1990)
723.
%%CITATION = NUPHA,B332,723;%%.

\bibitem{Grigoriev:2000rn}
M.~A. Grigoriev and S.~L. Lyakhovich, ``Fedosov deformation quantization as a
  {BRST} theory,'' {\em Commun. Math. Phys.} {\bf 218} (2001) 437--457,
\href{http://www.arXiv.org/abs/hep-th/0003114}{{\tt hep-th/0003114}}.
%%CITATION = HEP-TH 0003114;%%.

\bibitem{Batalin:2001je}
I.~A. Batalin, M.~A. Grigoriev, and S.~L. Lyakhovich, ``Star product for second
  class constraint systems from a {BRST} theory,'' {\em Theor. Math. Phys.}
  {\bf 128} (2001) 1109--1139,
\href{http://www.arXiv.org/abs/hep-th/0101089}{{\tt hep-th/0101089}}.
%%CITATION = HEP-TH 0101089;%%.

\bibitem{Bengtsson:1990un}
A.~K.~H. Bengtsson, ``{BRST} quantization in anti-de {S}itter space and gauge
  fields,'' {\em Nucl. Phys.} {\bf B333} (1990) 407.

\bibitem{Buchbinder:2001bs}
I.~L. Buchbinder, A.~Pashnev, and M.~Tsulaia, ``Lagrangian formulation of the
  massless higher integer spin fields in the {AdS} background,'' {\em Phys.
  Lett.} {\bf B523} (2001) 338--346,
\href{http://www.arXiv.org/abs/hep-th/0109067}{{\tt hep-th/0109067}}.
%%CITATION = HEP-TH 0109067;%%.

\bibitem{Sagnotti:2003qa}
A.~Sagnotti and M.~Tsulaia, ``On higher spins and the tensionless limit of
  string theory,'' {\em Nucl. Phys.} {\bf B682} (2004) 83--116,
\href{http://www.arXiv.org/abs/hep-th/0311257}{{\tt hep-th/0311257}}.
%%CITATION = HEP-TH 0311257;%%.

\bibitem{Lopatin:1988hz}
V.~E. Lopatin and M.~A. Vasiliev, ``Free massless bosonic fields of arbitrary
  spin in d- dimensional de {S}itter space,'' {\em Mod. Phys. Lett.} {\bf A3}
  (1988)
257.
%%CITATION = MPLAE,A3,257;%%.

\bibitem{Gaberdiel:1997ia}
M.~R. Gaberdiel and B.~Zwiebach, ``Tensor constructions of open string theories
  {I}: Foundations,'' {\em Nucl. Phys.} {\bf B505} (1997) 569--624,
\href{http://www.arXiv.org/abs/hep-th/9705038}{{\tt hep-th/9705038}}.
%%CITATION = HEP-TH 9705038;%%.

\bibitem{Barnich:2003wj}
G.~Barnich and M.~Grigoriev, ``Hamiltonian {BRST} and {B}atalin-{V}ilkovisky
  formalisms for second quantization of gauge theories,'' {\em Commun. Math.
  Phys.} {\bf 254} (2005) 581--601,
\href{http://www.arXiv.org/abs/hep-th/0310083}{{\tt hep-th/0310083}}.
%%CITATION = HEP-TH 0310083;%%.

\bibitem{Dresse:1990dj}
A.~Dresse, P.~Gr\'egoire, and M.~Henneaux, ``Path integral equivalence between
  the extended and nonextended {H}amiltonian formalisms,'' {\em Phys. Lett.}
  {\bf B245} (1990) 192.

\bibitem{Thorn:1987qj}
C.~B. Thorn, ``Perturbation theory for quantized string fields,'' {\em Nucl.
  Phys.} {\bf B287} (1987)
61.
%%CITATION = NUPHA,B287,61;%%.

\bibitem{Bochicchio:1987zj}
M.~Bochicchio, ``Gauge fixing for the field theory of the bosonic string,''
  {\em Phys. Lett.} {\bf B193} (1987)
31.
%%CITATION = PHLTA,B193,31;%%.

\bibitem{Bochicchio:1987bd}
M.~Bochicchio, ``String field theory in the {S}iegel gauge,'' {\em Phys. Lett.}
  {\bf B188} (1987)
330.
%%CITATION = PHLTA,B188,330;%%.

\bibitem{Barnich:1996mr}
G.~Barnich and M.~Henneaux, ``Isomorphisms between the {B}atalin-{V}ilkovisky
  antibracket and the {P}oisson bracket,'' {\em J. Math. Phys.} {\bf 37} (1996)
  5273--5296,
\href{http://www.arXiv.org/abs/hep-th/9601124}{{\tt hep-th/9601124}}.
%%CITATION = HEP-TH 9601124;%%.

\bibitem{Bonelli:2003zu}
G.~Bonelli, ``On the covariant quantization of tensionless bosonic strings in
  {AdS} spacetime,'' {\em JHEP} {\bf 11} (2003) 028,
\href{http://www.arXiv.org/abs/hep-th/0309222}{{\tt hep-th/0309222}}.
%%CITATION = HEP-TH 0309222;%%.

\bibitem{Barnich:2005ru}
G.~Barnich and M.~Grigoriev, ``{BRST} extension of the non-linear unfolded
  formalism,''
\href{http://www.arXiv.org/abs/hep-th/0504119}{{\tt hep-th/0504119}}.
%%CITATION = HEP-TH 0504119;%%.

\bibitem{Vasiliev:1994gr}
M.~A. Vasiliev, ``Unfolded representation for relativistic equations in (2+1)
  anti-{D}e {S}itter space,'' {\em Class. Quant. Grav.} {\bf 11} (1994)
649--664.
%%CITATION = CQGRD,11,649;%%.

\bibitem{Shaynkman:2000ts}
O.~V. Shaynkman and M.~A. Vasiliev, ``Scalar field in any dimension from the
  higher spin gauge theory perspective,'' {\em Theor. Math. Phys.} {\bf 123}
  (2000) 683--700,
\href{http://www.arXiv.org/abs/hep-th/0003123}{{\tt hep-th/0003123}}.
%%CITATION = HEP-TH 0003123;%%.

\bibitem{Barnich:2005bn}
G.~Barnich, N.~Bouatta, and M.~Grigoriev, ``Surface charges and dynamical
  {K}illing tensors for higher spin gauge fields in constant curvature
  spaces,'' {\em JHEP} {\bf 10} (2005) 010,
\href{http://www.arXiv.org/abs/hep-th/0507138}{{\tt hep-th/0507138}}.
%%CITATION = HEP-TH 0507138;%%.

\bibitem{Schneiders}
J.-P. Schneiders, ``An introduction to {$D$}-modules,'' {\em Bull. Soc. Royale
  Sci. Li{\`e}ge,} {\bf 63(3-4)} (1994) 223--295.

\end{thebibliography}
\end{document}